\documentclass[twocolumn,prb,showpacs,amsmath
  ,superscriptaddress]{revtex4}
\usepackage{epsfig,graphicx}

\def\be{\begin{equation}}
\def\ee{\end{equation}}
\def\e#1{\label{#1}\end{equation}}
\def\bea{\begin{eqnarray}}
\def\eea{\end{eqnarray}}
\def\ea#1{\label{#1}\end{eqnarray}}
\def\bes#1{\begin{subequations}\label{#1}}
\def\ese{\end{subequations}}

\begin{document}
%\twocolumn[\hsize\textwidth\columnwidth\hsize
%\csname@twocolumnfalse%
%\endcsname
%\draft
\title{Analysis of measurement errors for a superconducting phase qubit}
\author{Qin Zhang}
\author{Abraham G. Kofman}
\altaffiliation[Permanent address: ]{Department of Chemical Physics,
The Weizmann Institute of Science, Rehovot 76100, Israel}
 \affiliation{Department of Electrical Engineering, University of
California, Riverside, California 92521}
\author{John M. Martinis}
 \affiliation{Department of Physics, University of
California, Santa Barbara, California 93106}
\author{Alexander N. Korotkov}
 \affiliation{Department of Electrical Engineering, University of
California, Riverside, California 92521}
\date{\today}

\begin{abstract}
We analyze several mechanisms leading to errors in a course of measurement
of a superconducting flux-biased phase qubit. Insufficiently long
measurement pulse may lead to nonadiabatic transitions between qubit states
$|1\rangle$ and $|0\rangle$, before tunneling through a reduced barrier is
supposed to distinguish the qubit states. Finite (though large) ratio of
tunneling rates for these states leads to incomplete discrimination between
$|1\rangle$ and $|0\rangle$. Insufficiently fast energy relaxation after the
tunneling of state $|1\rangle$ may cause the repopulation of the quantum
well in which only the state $|0\rangle$ is supposed to remain. We analyze
these types of measurement errors using analytical approaches as well as
numerical solution of the time-dependent Schr\"{o}dinger equation.
 \end{abstract}

%\vskip1pc]

\maketitle

\section{Introduction}

In recent years, a significant progress has been made in developing
superconducting Josephson-junction circuits for quantum computation.
 Single-qubit \cite{charge-single,flux-single,phase-single} and coupled-qubit
\cite{charge-coupled,phase-coupled,mcd05} operations have been demonstrated
for various qubit types, including charge,
\cite{charge-single,charge-coupled} flux, \cite{flux-single} and phase
\cite{flux-single,phase-coupled,mcd05,coo04,sim04,joh05} qubits.
 Recently introduced flux-biased phase qubits
\cite{mcd05,coo04,sim04,joh05} (Fig.\ 1) significantly reduce the total
dissipation during measurement in comparison with the current-biased phase
qubits, \cite{phase-single,phase-coupled} which leads to lower decoherence
and faster recovery after measurement.

Measurement of qubits is an important stage in quantum information
processing.
 A novel scheme for fast measurements of flux-biased phase
qubits has been demonstrated \cite{coo04} and subsequently employed for
simultaneous measurement of two coupled qubits. \cite{mcd05}
 According to this scheme [see Fig.\ 1(b)], a measurement is performed by lowering the
barrier between the qubit ``left'' potential well and a much deeper ``right''
well, so that tunneling from qubit state $|1\rangle$ (the first excited
left-well state) to the right well occurs with probability close to one,
while state $|0\rangle$ (the left-well ground state) remains intact.

 It is of interest to analyze the measurement process theoretically,
since such an analysis can be used to minimize measurement errors.
 In measurements of coupled qubits, there are generally one- and
two-qubit errors.
 The main source of two-qubit errors is crosstalk, \cite{mcd05} arising
after the application of the measurement pulse due to the coupling
between the qubits.
 The crosstalk was analyzed by us recently. \cite{pre}

\begin{figure}
\centering
\includegraphics[width=3.3in]{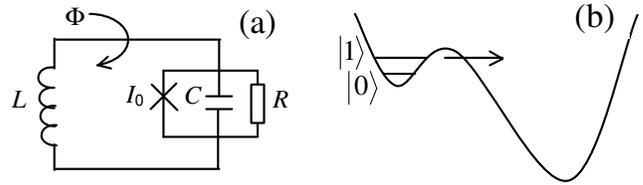}
\caption{(a) Circuit schematic of the flux-biased phase qubit, controlled by
external magnetic flux $\phi (t)$. (b) Sketch of the qubit potential energy
$U(\delta )$, where $\delta$ is the superconducting phase difference across
the Josephson junction. During the measurement pulse the energy barrier is
lowered so that the state $|1\rangle$ escapes from the ``left'' well into the
neighboring ``right'' well; this escape is sensed by a nearby SQUID.}
\label{Schematic}\end{figure}

The present paper is devoted to the theoretical study of the
behavior of a single flux-biased phase qubit during the measurement
pulse, which lowers the barrier.
 Our purpose is to find  potential sources of one-qubit
measurement errors in the fast measurement scheme used in Refs.\
\onlinecite{mcd05} and \onlinecite{coo04}, and analyze relations between the
error level and the parameters of the measurement pulse.
 We consider three mechanisms of single-qubit errors.
 First, there is a possibility of transitions between qubit states
$|1\rangle$ and $|0\rangle$  due to nonadiabaticity of the process.
 Second, there may be an incomplete discrimination between the two qubit states by
tunneling. Third, it happens that if the after-tunneling energy relaxation is
not fast enough, there is a possibility of a repopulation of the left qubit
well (in which only the state $|0\rangle$ is supposed to remain) after
tunneling of the state $|1\rangle$ to the right well.

The paper is organized as follows.
 Section \ref{II} contains a discussion of flux-biased phase qubits,
measurement errors, and the method for numerical solution of the
Schr\"{o}dinger equation used in the paper.
 Nonadiabatic effects are studied in Sec. \ref{III}, where we derive
analytical formulas and perform numerical calculations. The efficiency of
the qubit state discrimination by tunneling is discussed in Sec.\ IV.
   The repopulation error mechanism is discussed in Sec.\ V, where we perform
full numerical simulation of the quantum evolution in absence of energy
dissipation, and estimate the rate of dissipation required to eliminate the
repopulation effect.
 Finally, Sec. VI contains concluding remarks.

\section{Model}
\label{II}

The flux-biased phase qubit [Fig.\ 1(a)] has the same circuit schematic as
the rf SQUID. \cite{lik86}
 In its quantum description the wavefunction
$\Psi(\delta,t)$ evolves according to the Schr\"{o}dinger equation with the
Hamiltonian
 \be
H(t)=\frac{\hat{p}^2}{2m}+U(\delta,t),
 \e{4.1}
where $\delta$ is the superconducting phase difference across the Josephson
junction, $\hat{p}=-i\hbar (
\partial/\partial\delta )$ is the momentum operator, $m=[\Phi_0/(2\pi)]^2C$ is
the effective mass, $C$ is the junction capacitance, $\Phi_0=h/(2e)$ is the
flux quantum, $e$ is the electron charge, and the qubit potential is
 \be
U(\delta,t)=E_{J}\left\{\frac{[\delta-\phi(t)]^2}
{2\lambda}-\cos\delta\right\},
 \e{2.1}
where $E_{J}=I_{0}\Phi_0/(2\pi)$ is the Josephson energy, $\lambda=2\pi
I_{0}L/\Phi_0$ is the dimensionless inductance, $\phi(t)=2\pi\Phi(t)/\Phi_0$
is the dimensionless external magnetic flux, $\Phi(t)$ is the magnetic flux
in the loop, $I_{0}$ is the critical current, and $L$ is the inductance.
 We will assume that before the measurement pulse the right well of the
qubit potential  [see Fig.\ 1(b)] is much deeper than the left well (as in
Refs.\ \onlinecite{coo04,mcd05,sim04,joh05}).
 The qubit levels $|0\rangle$ and $|1\rangle$ are, respectively,
the ground and the first excited levels in the left well.

 The depth $\Delta U_{l,r}$ of a left ($l$) or right ($r$) well
(i.e.\ the energy difference between the potential maximum and minimum)
depends on external flux $\phi$ and can be characterized by the estimated
number of the discrete levels in the well:
 \be
 N_{l,r}=\frac{\Delta U_{l,r}}{\hbar\omega_{l,r}},
 \e{2.33}
where $\omega_{l,r}$ is the classical oscillation frequency near the well
bottom (the ``plasma frequency''), $\omega_{l,r}=\sqrt{E_J(1/\lambda +\cos
\delta_{l,r})/m}$ (here $\delta_{l,r}$ corresponds to the well bottom). The
energy dissipation in the phase qubit can be described by introducing
resistance $R$ into the circuit [Fig.\ \ref{Schematic}(a)]; however, for the
most of this paper the dissipation is neglected ($R=\infty$).

 The fast qubit state measurement\cite{coo04} is achieved by applying a
pulse of the magnetic flux $\phi$, which lowers sufficiently slowly
(adiabatically) the barrier height $\Delta U_l$ (and so $N_l$), until the
first excited state $|1\rangle$ of the qubit left well becomes very close to
the barrier top, so that the system in this state can easily tunnel through
the barrier to the right well, while tunneling from the ground state
$|0\rangle$ is almost negligible. (In principle, the measurement can be
arranged in a way so that the state $|1\rangle$ goes over the barrier;
however, in this case tunneling from the state $|0\rangle$ becomes
significant.)
 After the measurement pulse the potential $U(\delta )$ is returned to its initial
shape by decreasing $\phi$ back to its value before the measurement.
 Ideally, the system initially in the upper qubit state $|1\rangle$
 should switch  after the measurement
to the right well, while the qubit initially in the ground state $|0\rangle$
should remain in the left well.
 However, at least two kinds of errors are possible in this process of qubit
 measurement.
First, finite duration of the measurement pulse (which is supposed to be
rather fast) and corresponding nonadiabatic effects during increase of the
flux $\phi (t)$, may lead to transitions between levels $|1\rangle$ and
$|0\rangle$ (and other levels) before the tunneling starts. Second, during
the tunneling stage the qubit in the state $|1\rangle$ may erroneously
remain in the left well and/or state $|0\rangle$ may erroneously switch to
the right well. One reason for the second type of error is finite ratio of
tunneling rates for the states $|1\rangle$ and $|0\rangle$ (ideally, it
should be infinitely large). Another reason is the incomplete switching of
the state $|1\rangle$ due to repopulation of the left well from the right
well during or {\it after} the tunneling stage.
 These kinds of errors will be considered in the following sections.

 To solve numerically the time-dependent Schr\"{o}dinger
equation, we use the multi-projection approach of Ref.\ \onlinecite{Chen}.
In more detail, we divide the pulse duration $\tau$ into small time
increments $\Delta t$ and at each moment $t_k=k\Delta t\ (k=0,1,\dots)$ we
 find the eigenvalues $E_n^k$ and eigenfunctions
$\psi_n^k(\delta)$ for the Hamiltonian $H(t_k)$, using the Fourier grid
Hamiltonian method \cite{Marston} (equivalently, the periodic pseudospectral
\cite{for96} method).
 Then the wavefunction is computed as
 \be
\Psi(\delta,t)=\sum_n a_n^k\psi_n^k(\delta)\,
 e^{-iE_n^k(t-t_k)/\hbar}\ \
 (t_k\le t\le t_{k+1}),
 \e{4.2}
where
 \be
 a_n^k=\int_{-\infty}^\infty [\psi_n^k(\delta)]^* \,
 \Psi(\delta,t_k)\, d\delta
 \e{4.15}
 is calculated using the wavefunction $\Psi(\delta,t_k)$
at the end of the previous timestep. Starting the evolution from the initial
qubit level $i$, we calculate
   the evolving population of the level $n$ as
\begin{equation}
P_{ni}(t) =|a_n^k|^2 , \ \ \mbox{for} \ t_k\le t\le t_{k+1}.
    \label{P_qb}
    \end{equation}

\section{Nonadiabatic effects}
\label{III}

During the stage of rising flux $\phi$ (before the tunneling), the
population of the state $|0\rangle$ is supposed to remain unchanged. (Notice
that  excitation of state $|1\rangle$ to higher levels does not lead to the
measurement error.) However, since increase of $\phi$ is not infinitely
slow, there may be nonadiabatic transitions between levels $|0\rangle$ and
$|1\rangle$ (and/or higher levels) leading to measurement errors. Such
nonadiabatic errors are discussed in this section.

For numerical calculations we use the circuit parameters of Ref.\
\onlinecite{mcd05}:
 \be
C=700\ \mbox{fF},\ L=0.72\mbox{ nH},\ I_0=1.7\mbox{ mkA}.
 \e{2.16}
Also, we assume that during the first half of the measurement pulse the
left-well dimensionless barrier $N$ (here and below we omit the subscript
$l$) decreases from $N=5$ to $N=1.355$, which corresponds to $\phi$ varying
from 5.09 to 5.31 (for $N=1.355$ the state $|1\rangle$ is very close to the
barrier top\cite{N-difference}).

\begin{figure}
\centering
\includegraphics[width=2.5in]{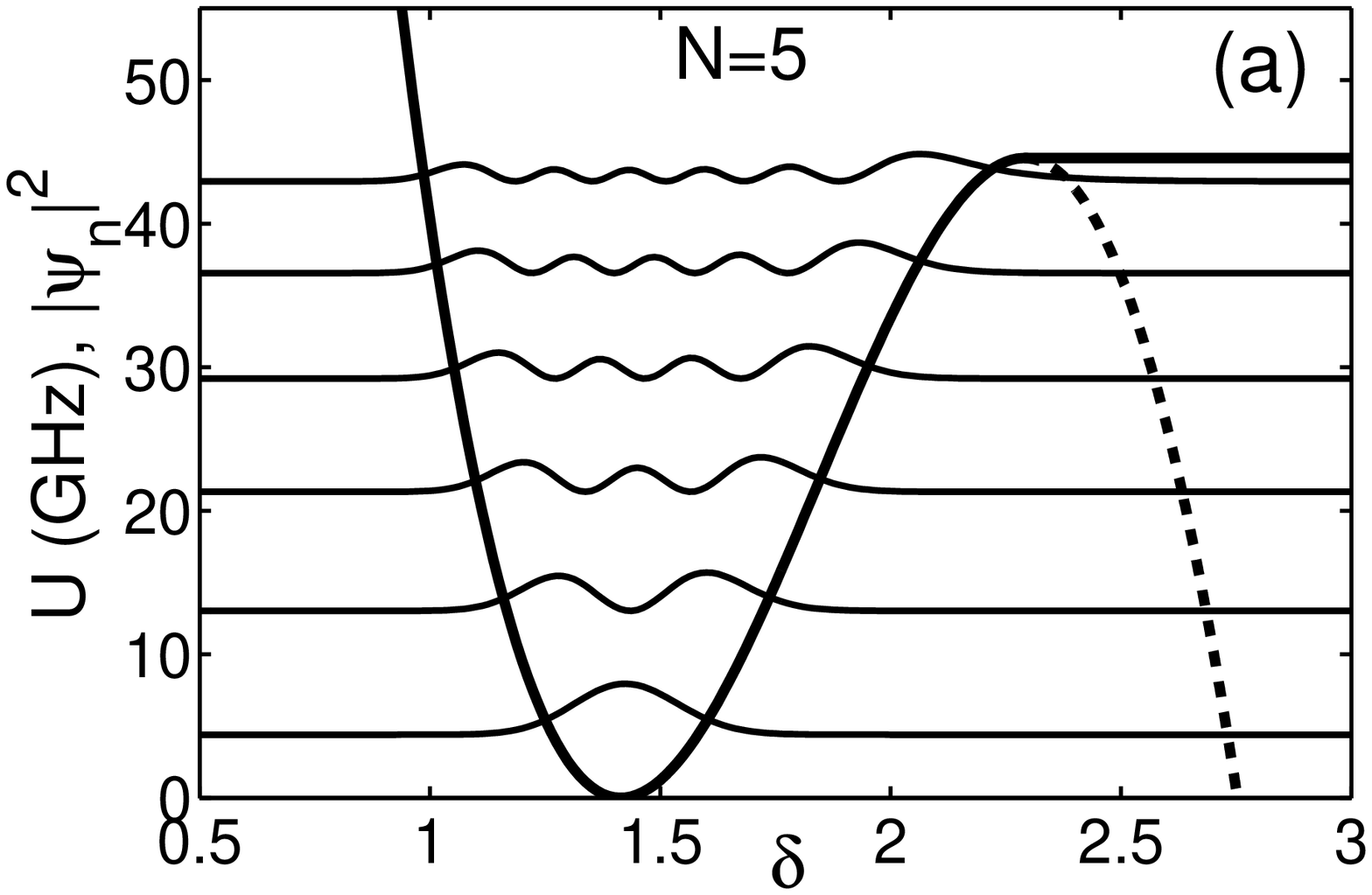}
\includegraphics[width=2.5in]{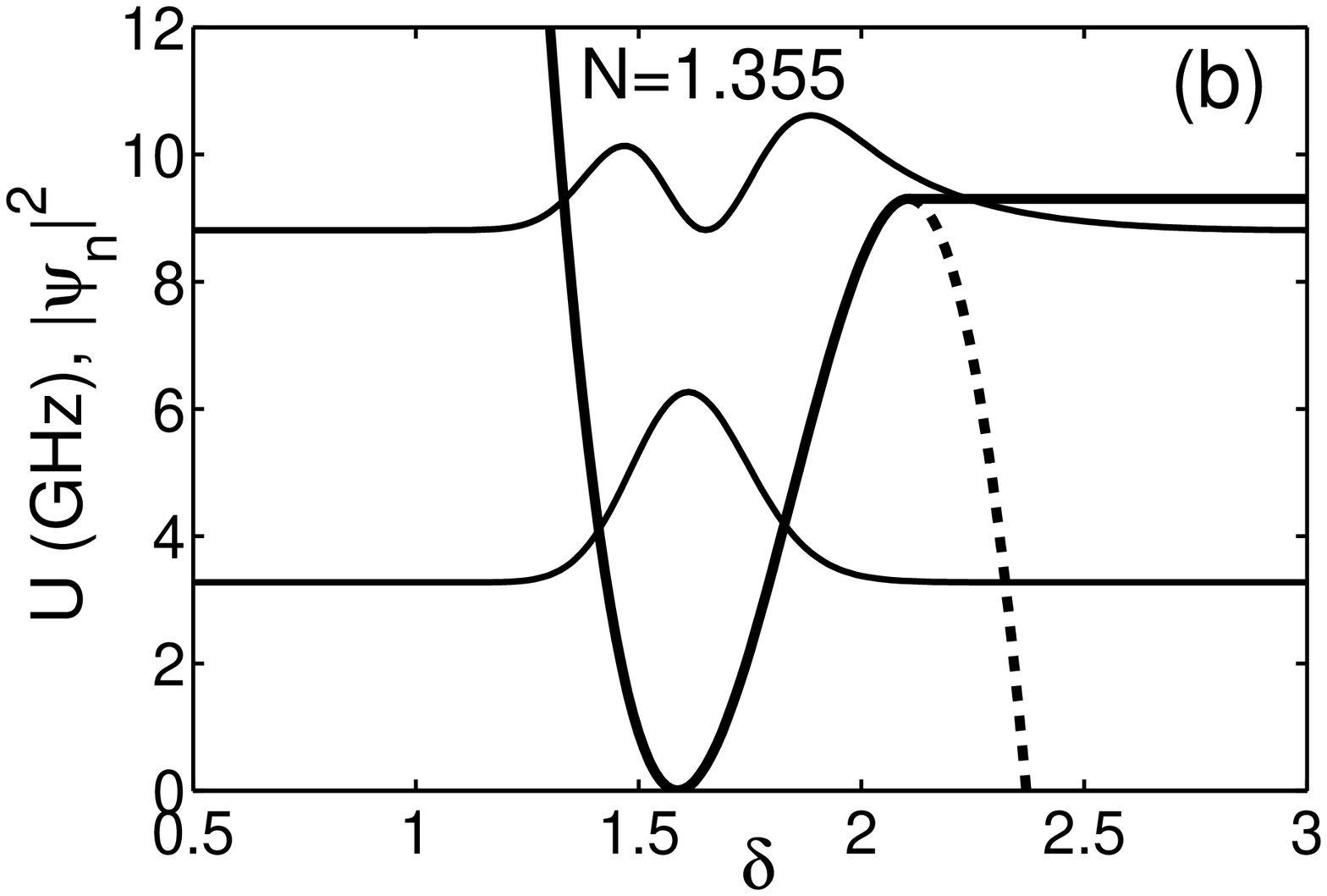}
%\vspace{0.4cm}
\caption{The modified qubit potential (\protect\ref{4.8}) (thick solid
lines, in GHz) and the normalized wavefunctions $|\psi_n(\delta)|^2$ (thin
solid lines), shifted vertically by the energy eigenvalues $E_n$ for (a)
$N=5$ and (b) $N=1.355$; other qubit parameters are given by
(\protect\ref{2.16}). Dashed lines correspond to the actual potential
(\protect\ref{2.1}).  The energy origin is chosen at the well minimum.}
\label{Potential}\end{figure}

For simplicity, in this section we neglect tunneling. Correspondingly, we
consider a modified potential energy (see Fig.\ \ref{Potential}), which
differs from Eq.\ (\ref{2.1}) by absence of the right well, so that the
potential is constant to the right of the barrier-top position $\delta_b$:
 \be
U(\delta,t)=\left\{
 \begin{array}{ll}
 E_{J}\{(\delta-\phi)^2/2\lambda-\cos\delta\},\ &
 \delta\le\delta_b, \vspace{0.2cm} \\
 U(\delta_b,t),&\delta>\delta_b,
 \end{array}\right.
 \e{4.8}
where both $\phi$ and $\delta_b$ change with time. In numerical calculations
we limit the range of $\delta$ between 0.5 and 3, so that the wavefunction
$\Psi (\delta, t)$ is assumed to vanish at $\delta=0.5$ and $\delta=3$ (we
have checked that further increase of this range does not change the
results). Figure \ref{Potential} shows the wavefunctions
$|\psi_n(\delta)|^2$ and energies $E_n$ for the states localized in the well,
corresponding to the beginning and the maximum of the measurement pulse.
Since the nonadiabatic transitions after the tunneling stage are not
important, in this section we consider only the first part of the
measurement pulse (increase of $\phi$, decrease of $N$).

\subsection{Analytical theory}

\subsubsection{Nonadiabatic error}

Assuming that the qubit wavefunction before the measurement is
 \be
|\Psi_0\rangle=c_0|\psi_0\rangle+c_1|\psi_1\rangle ,
 \e{3.2}
we describe the qubit evolution during the rising part of the measurement
pulse by expanding its wavefunction over the instantaneous basis given by
the eigenfunctions $\psi_n(\delta,t)$ of $H(t)$ with the eigenvalues
$E_n(t)$:
 \be
\Psi(\delta,t)=\sum_nc_n(t)e^{-(i/\hbar)\int_0^tE_n(t')dt'} \psi_n(\delta,t)
;
 \e{4.6}
the initial conditions are $c_{0}(0)=c_{0}$, $c_{1}(0)=c_{1}$, and $c_n(0)=0$
for $n\ge 2$.

    Let us define the nonadiabatic measurement error $Q$ as
the difference between the ground state occupation
$P_0(t)=|\langle\psi_0(t)|\Psi (t)\rangle|^2$ before the measurement pulse
($t=0$) and at the maximum of the pulse ($t=\tau$):
    \be
Q=|P_0(0)-P_0(\tau)|.
    \e{Q-def}
Here $P_0 (0) =|c_0|^2$, while $P_0(\tau )$ can be expressed via the
evolution operator $S(t)$ [so that $|\Psi(t)\rangle=S(t)|\Psi_0\rangle$]:
 \begin{eqnarray}
P_0(\tau )=|c_0|^2 P_{00}(\tau )+ |c_1|^2 P_{01}(\tau )
   \nonumber \\
 +2\, \mbox{Re}[c_0 c_1^* S_{00}(\tau )S_{01}^*(\tau )],
 \label{3.3}\end{eqnarray}
where $S_{ni}(\tau )=\langle\psi_n(\tau )|S(\tau )|\psi_i(0)\rangle$, while
$P_{ni}(\tau )=|S_{ni}(\tau )|^2$ is the occupation of state $n$ for initial
state $i$. The last term in the rhs of Eq.\ (\ref{3.3}) results from quantum
interference.

We are interested in the case of a small nonadiabatic error, $Q\ll 1$.
 Then, as will be shown below, $S_{01}(\tau )$ and $1-S_{00}(\tau )$ are
quantities of the first and second order in the perturbation, respectively.
 Hence, up to the second order, the nonadiabatic
error is given by
 \be
Q=|\, |c_0|^2[1-P_{00}(\tau )]-|c_1|^2 P_{01}(\tau )
 -2\,\mbox{Re}[c_0 c_1^* S_{01}^*(\tau)]|,
 \e{3.4}
Notice that $1-P_{00}$ and $P_{01}$ are both of the second order in
perturbation; therefore in the case when both qubit states are initially
occupied, the first two terms in Eq.\ (\ref{3.4}) can be neglected in
comparison with the third (interference) term. The third term is maximized
when both qubit states are equally occupied, $|c_0|=|c_1|=1/\sqrt{2}$, and
$c_0 c_1^* S_{01}^*(\tau )$ is a real number. So, the maximum nonadiabatic
error is approximately
   \be
 Q_e=|S_{01}(\tau )|=\sqrt{P_{01}(\tau )}.
  \e{3.5}

Notice that if initial state is $|\psi_1\rangle$, then the nonadiabatic
error is $Q=P_{01}(\tau )$. If initial state is $|\psi_0\rangle$, then the
error is $1-P_{00}(\tau )=\sum_{n\geq 1}P_{n0}(\tau )$, which as will be seen
later is approximately equal to $P_{01}(\tau )$ (because $P_{10}\approx
P_{01}$ in the lowest order, while $P_{n0}$ for $n\geq 2$ are very small
because direct transitions between non-neighboring levels are suppressed).
Therefore, if only one qubit state is initially occupied, the error (in the
lowest order) is
 \be
P_e = P_{01}(\tau)=Q_e^{\,2},
 \e{3.7}
which is much smaller than $Q_e$. This fact can be easily understood by
visualizing operator $S$ as a rotation of the qubit Bloch sphere (neglecting
higher levels). Then rotation by a small angle $\varphi$ (around $x$-axis)
leads to the change of $z$-coordinate by $\varphi$ near equator (for
$x=z=0$), while the change is only $\varphi^2/2$ near the poles.

    In this section we will mainly analyze the error $P_e$; however, it
should be remembered that the actual nonadiabatic error can be up to
$Q_e=\sqrt{P_e}$.

\subsubsection{Adiabatic perturbation theory.}
\label{IIIA2}

The evolution operator $S(t)$ obeys the Schr\"{o}dinger equation, which can
be reduced to the following form:\cite{hol92,got03}
%JQ.67.12;86.5
 \be
\dot{S}_{ni}=\sum_{m\ne n}
 \frac{\langle\psi_n(t)|\dot{H}(t)|\psi_m(t)\rangle}
{\hbar\omega_{nm}(t)}e^{i\int_0^t\omega_{nm}(t')dt'}S_{mi},
 \e{4.7}
where $\omega_{nm}=(E_n-E_m)/\hbar$. (The state label $i$ appears only in
indices, while $i$ in the exponent is the imaginary unit.)
 The initial condition for Eq.\ (\ref{4.7}) is $S_{ni}(0)=\delta^K_{ni}$,
where $\delta^K_{ni}$ is the Kronecker symbol.

 Equation (\ref{4.7}) is exact. It can be simplified assuming
slow variation of $H(t)$, so that $\dot{H}$ can be treated as a small
perturbation. In the zero approximation (the adiabatic approximation
\cite{hol92,got03}) the rhs of Eq.\ (\ref{4.7}) is assumed to vanish, so that
$\dot S=0$ and therefore $S_{ni}(t)=\delta^K_{ni}$. In the first
approximation we use $S_{mi}=\delta^K_{mi}$ in the rhs of Eq.\ (\ref{4.7}),
which yields \cite{got03} (for $n\ne i$)
 \be
S_{ni}(t)=\int_0^tdt'
 \frac{\langle\psi_n(t')|\dot{H}(t')|\psi_i(t')\rangle}
{\hbar\omega_{ni}(t')}e^{i\int_0^{t'}\omega_{ni}(t'')dt''}.
 \e{4.10}
Notice that for $n\ne i$, $S_{ni}(t)$ is of the first order in perturbation,
while $1-S_{ii}(t)$ vanishes in the first order [see Eq.\ (\ref{4.7})], so
that non-zero $1-S_{ii}(t)$ appears only in the second order, when the first
order $S_{mi}$ is used in Eq.\ (\ref{4.7}). Also notice that in the first
order $S_{ni}(t)=-S_{in}^*(t)$, as follows from Eq.\ (\ref{4.10}); therefore
$P_{ni}(t)=P_{in}(t)$. We have used these properties in the previous
subsection.

For the modified potential (\ref{4.8}) we find
 \be
\dot{H}(t)=\dot{U}(\delta,t)=\left\{
 \begin{array}{ll}
 E_J[\phi(t)-\delta]\dot{\phi}(t)/\lambda,&\delta\le\delta_b(t); \vspace{0.2cm} \\
 dU(\delta_b(t),t)/dt,&\delta>\delta_b(t).
 \end{array}\right.
 \e{4.9}
For simplicity let us use the formula in the first line of Eq.\ (\ref{4.9})
even at $\delta>\delta_b$ (this approximation can be justified by the fact
that the wavefunctions are small to the right of the barrier top position
$\delta_b$). Then the matrix elements of $\dot{H}$ can be calculated as
 \be
\langle\psi_n(t)|\dot{H}(t)|\psi_i(t)\rangle =
 -E_J\, \dot{\phi}(t)\,\delta_{ni}(t)/\lambda,
 \e{4.11}
 where
 \be
\delta_{ni}(t)=\int_{-\infty}^\infty\psi_n^*(\delta,t)\, \delta \,
\psi_{i}(\delta,t)\, d\delta
 \e{4.12}
is the ``position'' matrix element.
 Inserting Eq.\ (\ref{4.11}) into (\ref{4.10}), we obtain the following equation
for $P_{ni}(t)=|S_{ni}(t)|^2$ ($n\neq i$) in the lowest order of
perturbation:
 \be
P_{ni}(t) = \left(\frac{E_J}{\hbar\lambda}\right)^2
 \left|\int_0^tdt'\frac{\dot{\phi}(t')\, \delta_{ni}(t')}
{\omega_{ni}(t')}e^{i\int_0^{t'}\omega_{ni}(t'')dt''}\right|^2.
 \e{4.14}
Notice that for purely harmonic potential the matrix elements $\delta_{ni}$
are non-zero only for levels which are nearest neighbors. This means that
$P_{ni}$ for non-nearest neighbors would appear only in higher orders in
perturbation. Since the qubit potential is not purely harmonic, there will
be direct transitions between non-neighbors; however, these transition are
significantly weaker than for nearest neighbors.

When the measurement pulse duration is scaled without change of the pulse
shape, one can write the external flux as
 \be
\phi(t)\rightarrow\phi(t/\tau)=\phi(\theta),
 \e{4.18}
where $\tau$ is the duration of the rising part of the pulse and
$\theta=t/\tau$. Then the Hamiltonian and hence its eigenstates and
eigenvalues depend only on $\theta$:
 \be
H(t)\rightarrow H(\theta),\ \
\omega_{ni}(t)\rightarrow\omega_{ni}(\theta),\ \
 \psi_n(t)\rightarrow\psi_n(\theta).
 \e{4.19}
As a result, Eq.\ (\ref{4.14}) can be recast in the form
 \be
P_{ni}(t)= \left(\frac{E_J}{\hbar\lambda}\right)^2
 \left|\int_0^\theta d\theta'\frac{\phi'(\theta')
\delta_{ni}(\theta')}{\omega_{ni}(\theta')}
e^{i\tau\int_0^{\theta'}\omega_{ni}(\theta'')d\theta''}\right|^2,
 \e{4.29}
where $\phi'(\theta)=d\phi(\theta)/d\theta$.

  Finally, the nonadiabatic measurement error defined as $P_e=P_{01}(\tau )$ [see Eq.\
(\ref{3.7})] is obtained as
   \be
P_e  = \left(\frac{E_J}{\hbar\lambda}\right)^2
 \left|\int_0^1d\theta\frac{\phi'(\theta)\delta_{10}(\theta)}
{\omega_{10}(\theta)}
e^{i\tau\int_0^{\theta}\omega_{10}(\theta')d\theta'}\right|^2.
 \e{Exact}
Even though this is only the lowest order approximation for $P_{01}(\tau )$,
the numerical calculations of $P_{01}$ show (see below) that Eq.\
(\ref{Exact}) is quite accurate.

     This formula for $P_e$ can be further simplified (though with a noticeable
loss of accuracy) by considering the qubit as a harmonic oscillator with the
frequency $\omega_{10}$ changing in time. In this approximation \cite{lan77}
 \be
 \delta_{10} = \sqrt{\hbar/(2m\omega_{10})},
 \e{4.23}
so Eq.\ (\ref{Exact}) becomes
 \be
P_e  = \frac{E_J^2}{2\hbar\lambda^2m}
 \left|\int_0^1d\theta\frac{\phi'(\theta)}
{\omega_{10}^{3/2}(\theta)}
e^{i\tau\int_0^{\theta}\omega_{10}(\theta')d\theta'}\right|^2.
 \e{Simple}
This expression requires only the knowledge of the instantaneous transition
frequency $\omega_{10}(t/\tau)$ and does not depend on the instantaneous
wavefunctions.

    Even further simplification is possible if we neglect the change
of the frequency $\omega_{10}$ with time (as we see later, this is a quite
crude approximation). Then the measurement error $P_e$ is simply determined
by the spectral component of the derivative $\phi '$ of the external flux at
the frequency $\omega_{10}$:
  \be
P_e  = \frac{E_J^2}{2\hbar\lambda^2m \omega_{10}^3}
 \left|\int_0^1 \phi'(\theta)\,
e^{i\omega_{10} \tau \theta} d\theta \right|^2.
  \e{simple-simple}

Since $\omega_{10}$ changes during the measurement pulse quite significantly
(by more than 50 \%, see Fig.\ \ref{Potential}), application of Eq.\
(\ref{simple-simple}) is not straightforward and requires a choice of the
time moment within the pulse, at which $\omega_{10}$ is taken. However, Eq.\
(\ref{simple-simple}) can still be used for simple estimates of the
nonadiabatic error, for example using the maximum and minimum values of
$\omega_{10}$ (at the beginning of the pulse and at the end of its rising
part). Notice that Eq.\ (\ref{simple-simple}) implies that the nonadiabatic
error can be significantly reduced by suppressing the spectral components of
$\dot{\phi}(t)$ in the frequency range of $\omega_{10}$.

\subsubsection{Long-$\tau$ behavior.}

In this subsection we analyze the asymptotic behaviour of the nonadiabatic
measurement error $P_e$ for long pulse duration $\tau$, assuming that the
pulse profile $\phi (t)$ is analytic everywhere except for the endpoints
$t=0$ and $t=\tau$. We show that in this case the long-$\tau$ behavior of
$P_e$ can be expressed via pulse properties at the singular (nonanalytic)
endpoints.\cite{gar62} Even though the result obtained below in this
subsection cannot be directly applied to a realistic experiment, it still
sheds a light on the dependence of the measurement error on the pulse shape.

 The $j$-fold integration by parts of Eq.\ (\ref{4.10}) yields
 %JQ.88.8
 \bea
&&S_{ni}(\tau)=-
 \frac{e^{i\int_0^{t}\omega_{ni}(t')dt'}}{\omega_{ni}(t)}
\sum_{k=1}^{j}i^{k}\left.
 \frac{d^{k-1}}{dt^{k-1}}\left[\frac{g(t)}{\omega_{ni}^{k-1}(t)}
 \right]\right|_0^\tau\nonumber\\
 &&+i^j\int_0^\tau e^{i\int_0^{t}\omega_{ni}(t')dt'}
 \frac{d^j}{dt^j}\left[\frac{g(t)}{\omega_{ni}^j(t)}\right]\, dt,
 \ea{4.16}
where $g(t)=\langle\psi_n(t)|\dot{H}(t)|\psi_i(t)\rangle /
\hbar\omega_{ni}(t)$ and the level label $i$ again appears only in indices
($i$ in other places is the imaginary unit). For scaling the pulse duration
$\tau$ without change of a pulse shape, we use relation
$d/dt=\tau^{-1}d/d\theta$ [see Eqs.\ (\ref{4.18}) and (\ref{4.19})]; then the
$k$th term in Eq.\ (\ref{4.16}) acquires a factor $\tau^{-k}$. Let us denote
by $j_0$ and $j_1$ the lowest orders of the time derivatives of the
Hamiltonian, which do not vanish at $t=0$ and $\tau$, respectively.
 Then for sufficiently large $\tau$,
$(\min \omega_{ni})\tau \gg 1$, one can leave in Eq.\ (\ref{4.16}) only the
lowest-order nonvanishing terms corresponding to $t=0$ and $\tau$
($\theta$=0 and 1) and obtain the approximation
 %JQ.89.1
 \bea
&&S_{ni}(\tau)\approx
 \frac{i^{j_0}\langle\psi_n(0)|H^{(j_0)}(0)|\psi_i(0)\rangle}
{\hbar\omega_{ni}^{j_0+1}(0)\tau^{j_0}}\nonumber\\
 &&-\frac{i^{j_1}\langle\psi_n(1)|H^{(j_1)}(1)|
\psi_i(1)\rangle}{\hbar\omega_{ni}^{j_1+1}(1)\tau^{j_1}}
 e^{i\bar{\omega}_{ni}\tau},
 \ea{4.17}
where time is normalized by $\tau$, $H^{(j)}(\theta)$ is the $j$th
derivative of $H(\theta)$, and
$\bar{\omega}_{ni}=\int_0^1\omega_{ni}(\theta)d\theta$ is the frequency
averaged over the pulse duration.

    Using the relation
$\langle\psi_n(\theta)|H^{(j)}(\theta)|\psi_i(\theta)\rangle =
 -E_J\phi^{(j)}(\theta)\, \delta_{ni}(\theta)/\lambda$,
which is similar to Eq.\ (\ref{4.11}), we obtain the following approximation
for the nonadiabatic error $P_e=|S_{01}(\tau )|^2$:
 \bea
P_e\approx\frac{E_J^2}{\hbar^2\lambda^2\tau^{2j}}
 \left|\frac{\phi^{(j_0)}(0)\delta_{10}(0)}
{\omega_{10}^{j_0+1}(0)\tau^{j_0-j}}\right.\nonumber\\
 -\left.\frac{i^{j_1-j_0}\phi^{(j_1)}(1)\delta_{10}(1)}
{\omega_{10}^{j_1+1}(1)\tau^{j_1-j}}
e^{i\bar{\omega}_{10}\tau}\right|^2,
 \ea{4.21}
where $j=\min\{j_0,j_1\}$.
 The error probability (\ref{4.21}) generally decays with $\tau$ as
 $\tau^{-2j}$.
 In particular, in the case $j_0=j_1=j$, Eq.\ (\ref{4.21}) becomes
 \be
P_e\approx\frac{E_J^2}{\hbar^2\lambda^2\tau^{2j}}
 \left|\frac{\phi^{(j)}(0)\delta_{10}(0)}{\omega_{10}^{j+1}(0)}
 -\frac{\phi^{(j)}(1)\delta_{10}(1)}
{\omega_{10}^{j+1}(1)}e^{i\bar{\omega}_{10}\tau}\right|^2.
 \e{4.22}
In this case besides decaying with $\tau$, the dependence $P_e(\tau )$ also
oscillates with the frequency $\bar{\omega}_{10}$.
 In contrast, when $j_0\ne j_1$, one of the two terms in Eq.\ (\ref{4.21})
dominates at large $\tau$, and the decay occurs monotonously.

\subsubsection{Time-dependent perturbation theory}

 The approach used in subsection \ref{IIIA2} can be compared with the standard
time-dependent perturbation theory. Let us split the Hamiltonian (\ref{4.1})
into the ``unperturbed'' Hamiltonian $H_0$ and ``perturbation'' $V(t)$:
 \begin{eqnarray}
&& H(t)=H_0+V(t),\ H_0=\frac{\hat{p}^2}{2m}+U_0(\delta),
 \label{3.10} \\
&& \hspace{0.5cm} U_0(\delta)=E_{J}\left[\frac{(\delta-\phi_r)^2}
{2\lambda}-\cos\delta\right],
 \label{3.11} \\
&& \hspace{0.5cm} V(t)=-(E_J/\lambda)[\phi(t)-\phi_r],
 \label{3.12}
 \end{eqnarray}
where we neglect the change of the energy origin and choose an arbitrary
flux $\phi_r$ as a reference point. Even though the standard time-dependent
perturbation theory \cite{lan77} can be only applied when the perturbation
$V(t)$ is small, which is obviously not the case in our problem, let us
still apply it formally. Then for the nonadiabatic error we obtain
 \bea
&&P_e =\left(\frac{E_J\delta_{10}} {\hbar\lambda\omega_{10}}\right)^2
 \left|\int_0^\tau \dot{\phi}(t)e^{i\omega_{10}t} dt \right|^2
\nonumber\\
&& =\left(\frac{E_J\delta_{10}} {\hbar\lambda\omega_{10}}\right)^2
 \left|\int_0^1 \phi'(\theta)
e^{i\omega_{10}\tau\theta} d\theta \right|^2,
 \ea{3.9}
where $\delta_{10}$ and $\omega_{10}$ correspond to the Hamiltonian $H_0$.
It is easy to see that Eq.\ (\ref{3.9}) coincides with Eq.\ (\ref{Exact}) if
the time-dependence of $\delta_{10}$ and $\omega_{10}$ in Eq.\ (\ref{Exact})
is neglected. An arbitrary choice of $\phi_r$ in the unperturbed Hamiltonian
with a natural limitation $\phi (0)\le \phi_r \le \phi (\tau) $ corresponds
to an arbitrary choice of the time moment within the pulse, at which
$\delta_{10}$ and $\omega_{10}$ are taken.

 Equation (\ref{3.9}) can be simplified by using the
approximation (\ref{4.23}). Then it becomes exactly Eq.\
(\ref{simple-simple}), which expresses $P_e$ via the spectral component of
the derivative of the external flux at the transition frequency
$\omega_{10}$. Even though this simplest result can be obtained in both
approaches, we would like to emphasize that the approach of subsection
\ref{IIIA2} allows us to use much more accurate approximations for the error
probability $P_e$.

\subsection{Numerical results}
 \label{IIIB}

For numerical calculations we use the method described in Section II [see
Eqs.\ (\ref{4.2})--(\ref{P_qb})]; which corresponds to the solution of exact
equation (\ref{4.7}).
 Let us describe the time dependence of the flux during the rising part
of the measurement pulse as
 \be
\phi(t/\tau)=\phi_0+(\phi_1-\phi_0)\times g(t/\tau),
 \e{4.24}
where $\phi_0$ and  $\phi_1$ are the initial and final flux values, while
$g(\theta)$ describes the pulse shape and satisfies conditions  $g(0)=0$ and
$g(1)=1$. Unless mention otherwise, we choose in this subsection
$\phi_0=5.09$ (corresponding to $N=5$) and $\phi_1=5.31$ (corresponding to
$N=1.355$).

\begin{figure}[tb]
\centering
\includegraphics[width=2.6in]{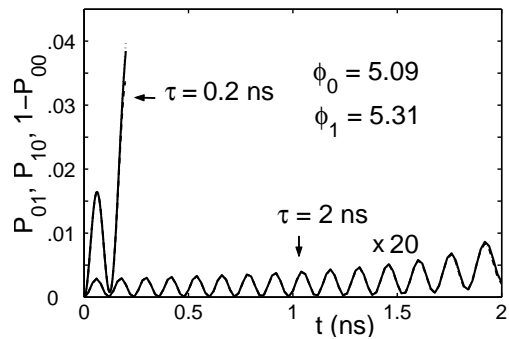}
%\vspace{.5cm}
 \caption{Time dependence of the instantaneous nonadiabatic error defined in
three different ways: $P_{01}(t)$ (solid lines), $P_{10}(t)$ (dashed lines),
and $1-P_{00}(t)$ (dotted lines), for two durations of the linear pulse:
$\tau=0.2$ ns and $\tau=2$ ns (for $\tau=2$ ns the data are multiplied by
the factor 20). The solid, dashed, and dotted lines are practically
indistinguishable, except near the pulse end for $\tau=0.2$ ns.
 }
 \label{P_0}\end{figure}

Consider first the simplest case of a linearly increasing pulse,
 $g(\theta)=\theta$.
 Figure \ref{P_0} shows the time evolution of
three quantities: $P_{01}$, $P_{10}$, and $1-P_{00}$ (which all could in
principle be used for the definition of the nonadiabatic error) for two
values of the pulse duration: $\tau=0.2$ ns and $\tau=2$ ns. One can see that
as expected, these three quantities practically coincide:
  \be
  P_{01} (t) \approx P_{10} (t) \approx 1-P_{00}(t),
  \e{p01=p10=1-p00}
so that it is not really important which of them is used for the definition
of the error $P_e$. (In this subsection for calculation of $P_e$ we will use
$1-P_{00}$, which is slightly larger than two other quantities.) The
dominating process is transitions between states $|0\rangle$ and
$|1\rangle$; as a result, the oscillations in Fig.\ \ref{P_0} are at the
frequency $\omega_{10}(t)$ [see Eq.\ (\ref{4.14})].
 The oscillation amplitude increases with time in Fig.\ \ref{P_0}.
 This is explained by the fact that increasing flux during the measurement
pulse decreases the transition frequency $\omega_{10}$ (see Fig.\
\ref{Potential}), which leads to the increase of the integrand in Eq.\
(\ref{4.14}) for $P_{01}(t)$ [note also Eq.\ (\ref{4.23})]; therefore, the
oscillation amplitude should increase approximately as
$[\omega_{10}(t)]^{-3}$.
 Figure \ref{P_0} also demonstrates that the instantaneous nonadiabatic error
is much larger for $\tau=0.2$ ns than for $\tau=2$ ns.

\begin{figure}[htb]
\centering
\hspace{-0.6cm}\includegraphics[width=2.65in]{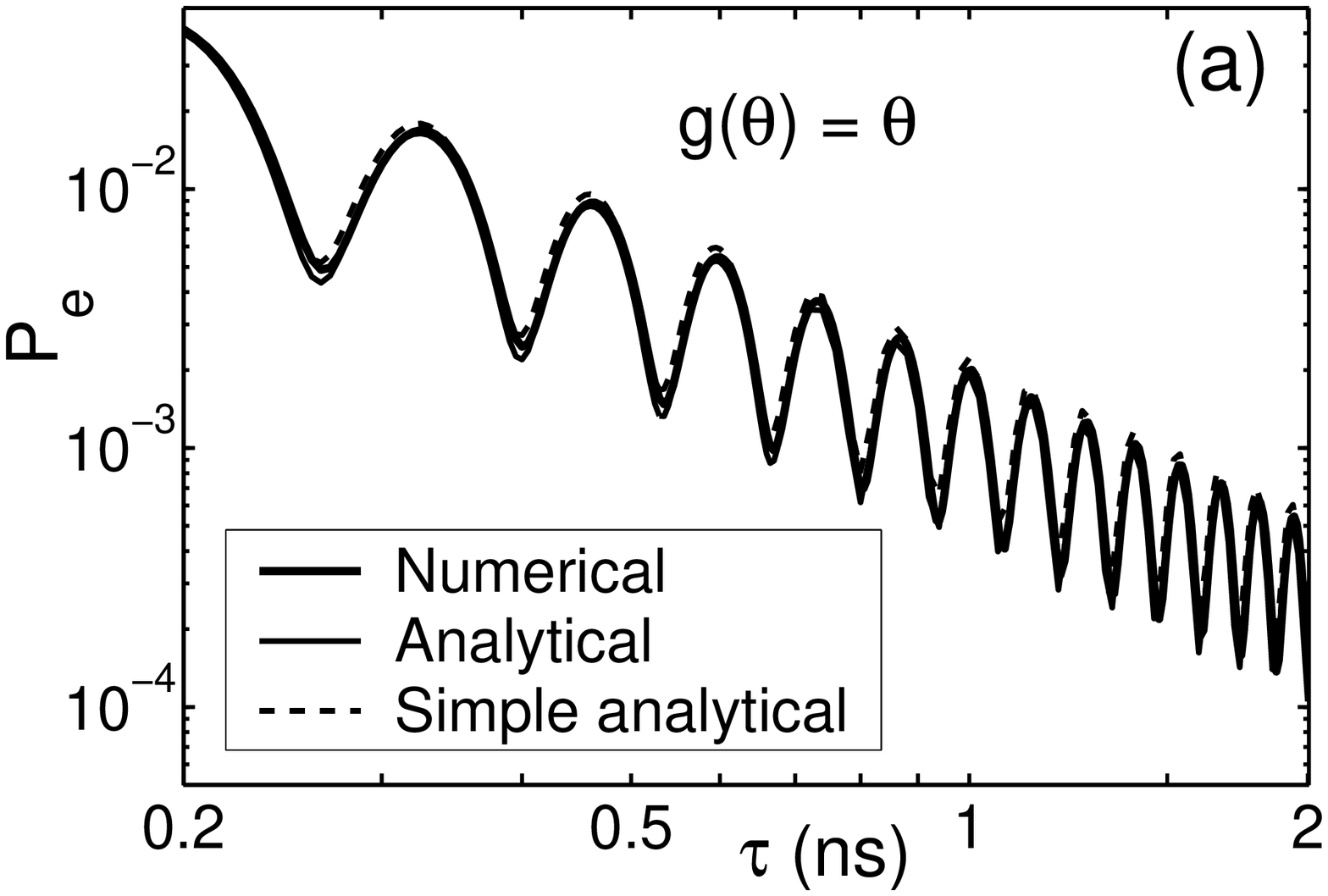}
\includegraphics[width=2.91in]{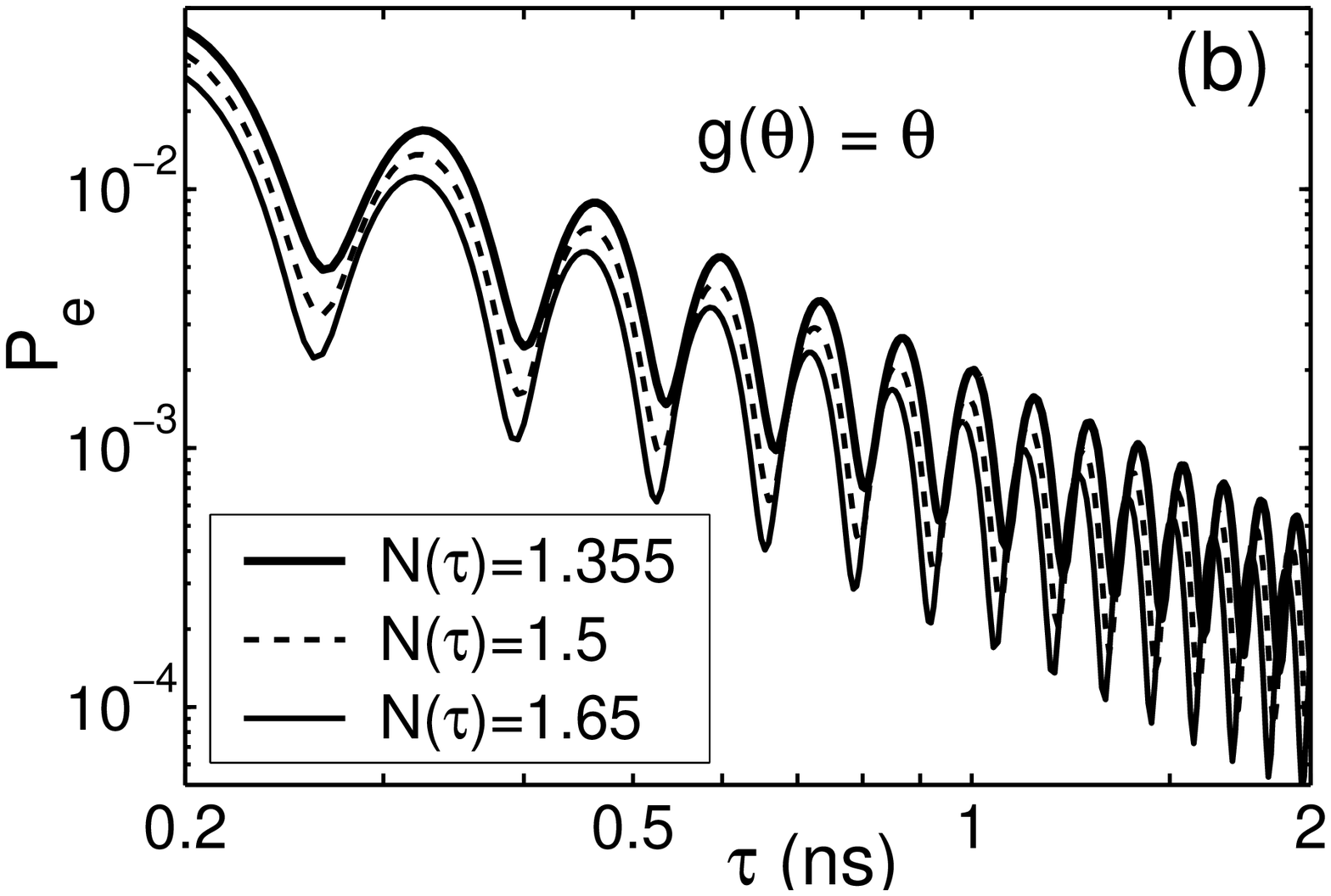}
  \caption{The nonadiabatic
error $P_e$ as a function of the duration $\tau$ of the linear measurement
pulse. (a) Numerical results (thick solid line) and results obtained using
Eq.\ (\ref{Exact}) (thin solid line) and Eq.\ (\ref{Simple}) (dashed line)
for the pulse starting with the barrier height $N=5$ (corresponding to
$\phi_0=5.09$) and ending with $N=1.355$ ($\phi_1=5.308$). (b) Numerical
results for pulses starting with $N=5$ and ending with $N=1.355$ (thick
solid line), $1.5$ (dashed line), and $1.65$ (thin solid line);
corresponding maximum fluxes are $\phi =5.308$, 5.298, and 5.287.
 }
 \label{f19}\end{figure}

    Figure \ref{f19} shows the dependence of the error $P_e$ on the duration
     $\tau$ of the linear pulse on the
double-logarithmic scale. The nonadiabatic error generally decreases with
increase of $\tau$ (for linear pulse crudely as $\tau^{-2}$); however, the
dependence also shows oscillations, which originate from oscillations in
Fig.\ \ref{P_0}.
 In Fig.\ \ref{f19}(a) the thick solid line shows the numerical
result. It is practically indistinguishable from the thin solid line showing
the lowest-order analytical result (\ref{Exact}). The dashed line
corresponds to the simplified formula (\ref{Simple}); one can see that it is
also quite close to the exact result. We have observed a similar relation
between the results obtained numerically, by analytical formula
(\ref{Exact}), and by simplified formula (\ref{Simple}) for other pulse
shapes discussed below.
 Figure \ref{f19}(b) shows the dependence of the error on the
pulse duration for three values of the final barrier height: $N =1.355$,
$1.5$ and $1.65$ (the initial value is $N=5$ for all the cases).
 One can see that the error increases with the decrease of the final
barrier height; however, dependence is not strong and therefore the results
presented here are not quite sensitive to the exact choice of the pulse
amplitude.

\begin{figure}[tb]
\centering
\includegraphics[width=2.65in]{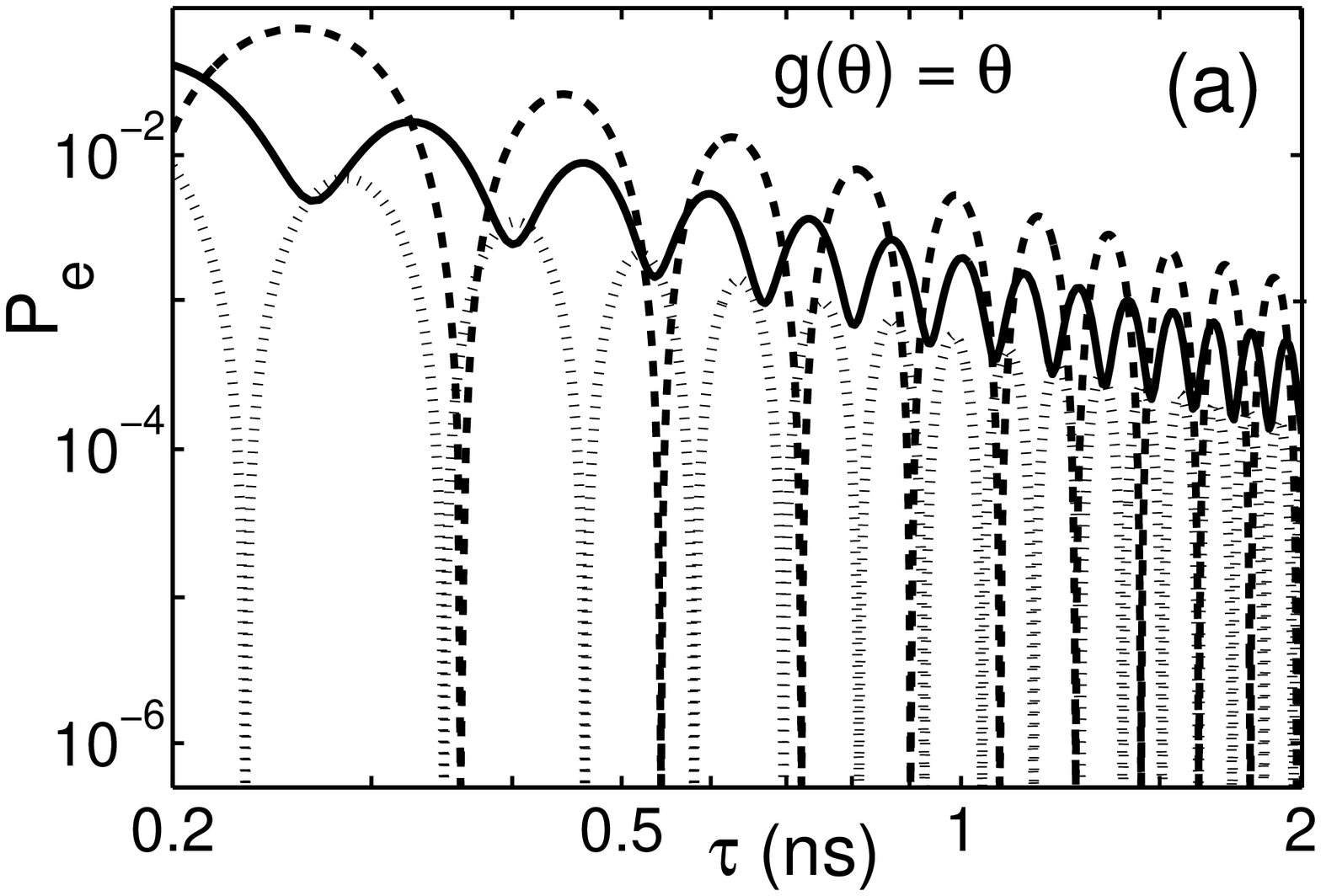}
\includegraphics[width=2.65in]{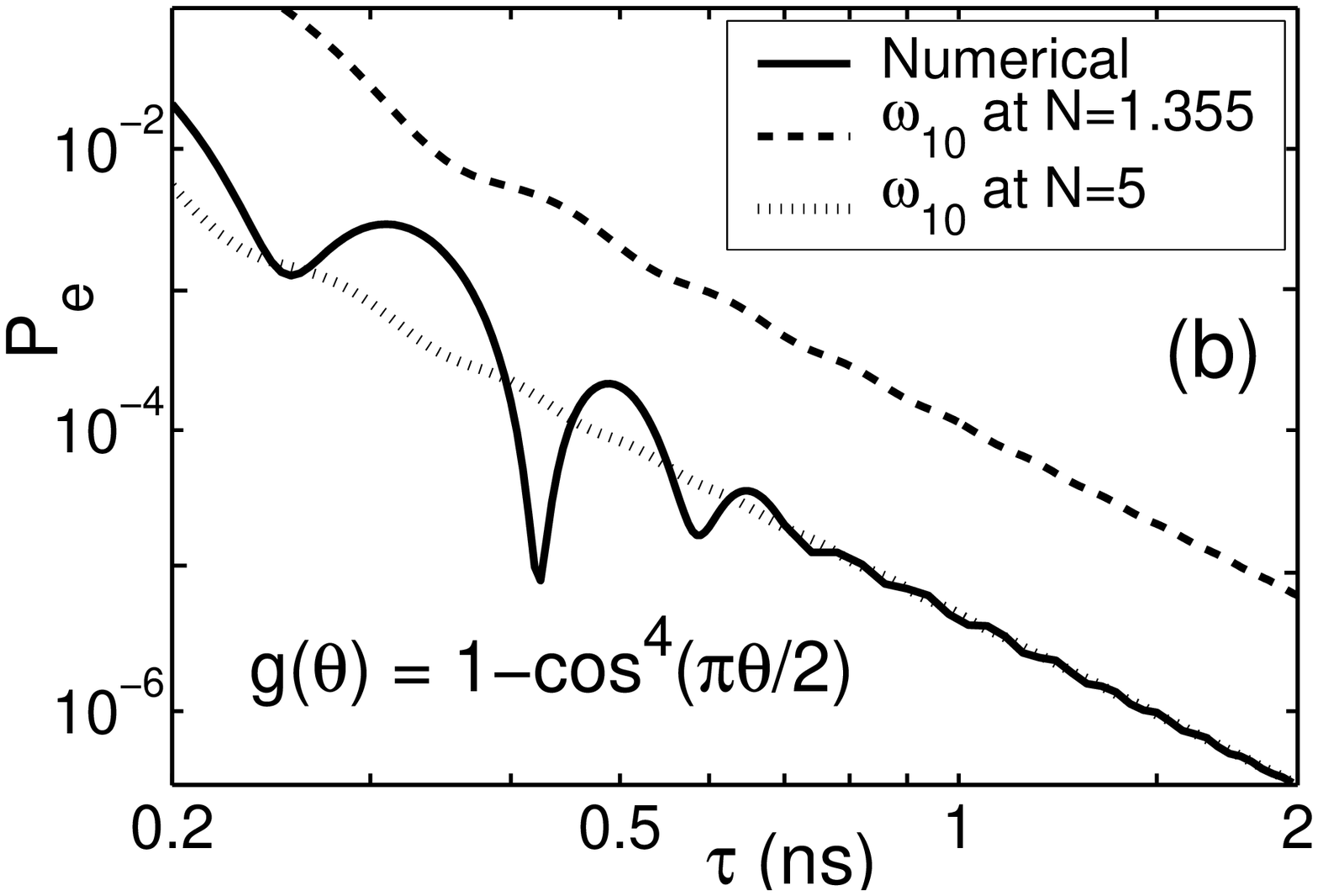}
\includegraphics[width=2.65in]{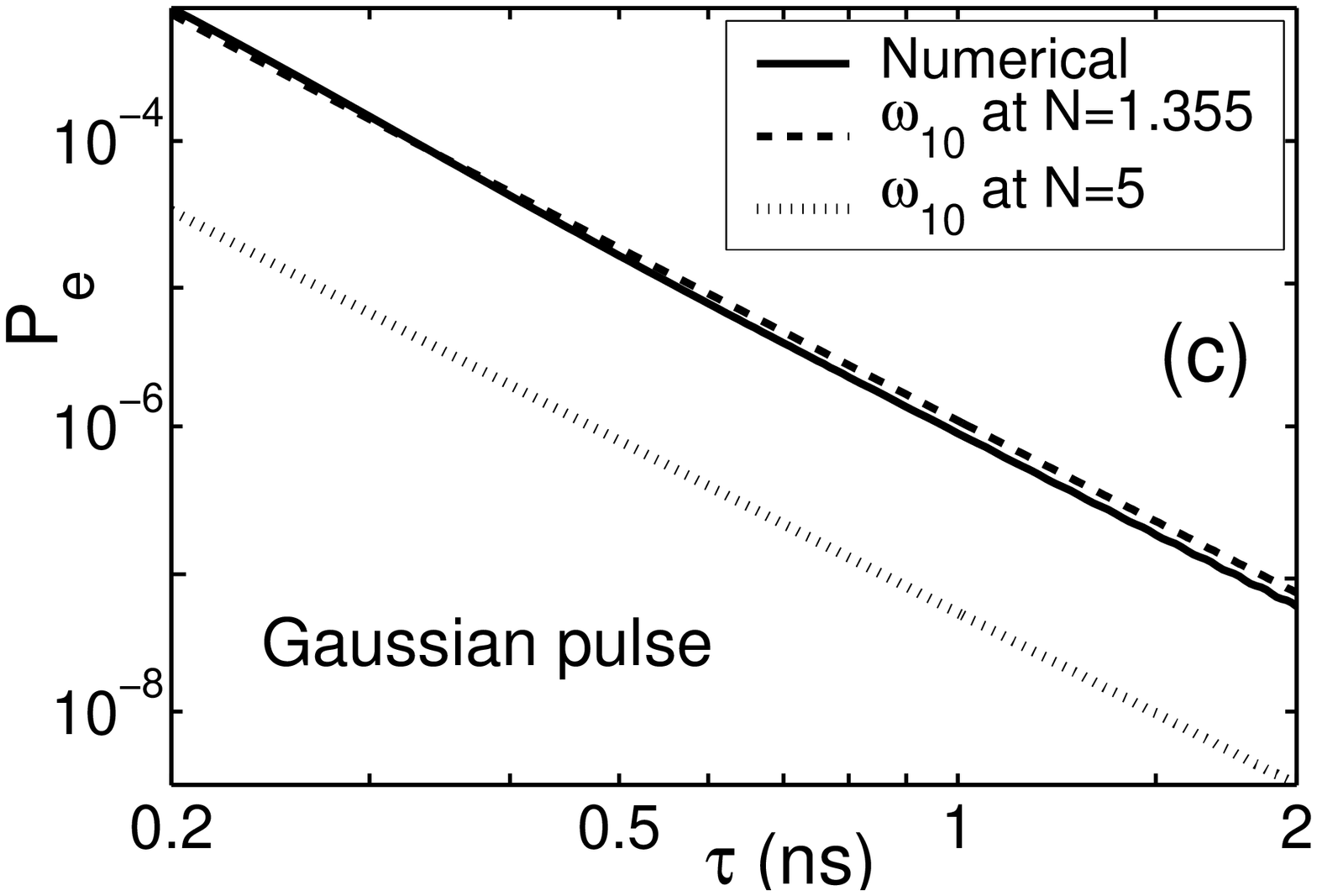}
\caption{Dependence of the measurement error $P_e$ on the pulse duration
$\tau$ calculated either numerically (solid lines) or using the simple
formula (\ref{simple-simple}), in which the spectral component is taken at
the frequency $\omega_{10}$ corresponding to either beginning (dotted lines)
or end (dashed lines) of the measurement pulse. The pulse is (a) linear,
$g(\theta)=\theta$, (b) of the shape $g(\theta )=1-\cos^4(\pi\theta /2)$, or
(c) of Gaussian shape. For the Gaussian pulse $\tau$ is defined as the
r.m.s. width of the pulse.
 }
 \label{comp-simple}\end{figure}

  Figure \ref{comp-simple} shows the comparison between the numerical results
for $P_e(\tau )$ dependence and the results obtained using Eq.\
(\ref{simple-simple}) which relates $P_e$ to the spectral component of
$\dot{\phi}$ at the frequency $\omega_{10}$. Three panels of Fig.\
\ref{comp-simple} are for (a) the linear pulse, (b) the pulse with shape
$g(\theta)=1-\cos^4(\pi\theta /2)$, and (c) Gaussian pulse. (For Gaussian
pulse we define $\tau$ as the r.m.s. width of the pulse; the fact that the
pulse is really longer than $\tau$, reduces $P_e$ in comparison with other
shapes.) The frequency $\omega_{10}$ in Eq.\ (\ref{simple-simple}) is not
well-defined since it changes during the pulse by more than 50\%; so in Fig.\
\ref{comp-simple} we show two analytical results, corresponding to
$\omega_{10}$ at the beginning and at the end of the rising part of the
pulse. One can see that in Fig.\ \ref{comp-simple}(a) the analytical curves
show oscillations as well as the numerical curve; however, the amplitude of
oscillations is much larger for the analytics ($P_e$ goes to zero at some
values of $\tau$, in contrast to the numerical results). Neglecting
oscillations, we see that the numerical curve is approximately in the middle
between the two analytical curves (on the logarithmic scale). In contrast to
that, in Fig.\ \ref{comp-simple}(b) the numerical curve is close to the
lower analytical curve (and shows some oscillations, which are not well
pronounced in the analytical results), while in Fig.\ \ref{comp-simple}(c)
the numerical curve is close to the upper analytical line. As follows from
the results in Fig.\ \ref{comp-simple}, even though Eq.\
(\ref{simple-simple}) can be used for a simple estimate of the nonadiabatic
error, the accuracy of this estimate can easily be worse than one order of
magnitude. [The ratio between the upper and lower analytical estimates in
Figs.\ \ref{comp-simple}(b) and \ref{comp-simple}(c)  is crudely
$r_\omega^7$, where $r_\omega$ is the ratio of frequencies $\omega_{10}$
before and in the middle of the pulse; and even though $r_\omega \simeq 1.6$
is not a big number, $r_\omega^7$ is already more than an order of
magnitude.]
 Notice that $P_e$ for the
Gaussian pulse [Fig.\ \ref{comp-simple}(c)] scales with $\tau$ as
$\tau^{-4}$, and not in a Gaussian way, as might be naively expected since
Fourier transform of a Gaussian is a Gaussian. The reason is that in our case
we should use only the rising half of the Gaussian pulse, for which Fourier
transform scales with frequency only as $\omega^{-2}$.

\begin{figure}
\centering
\includegraphics[width=2.65in]{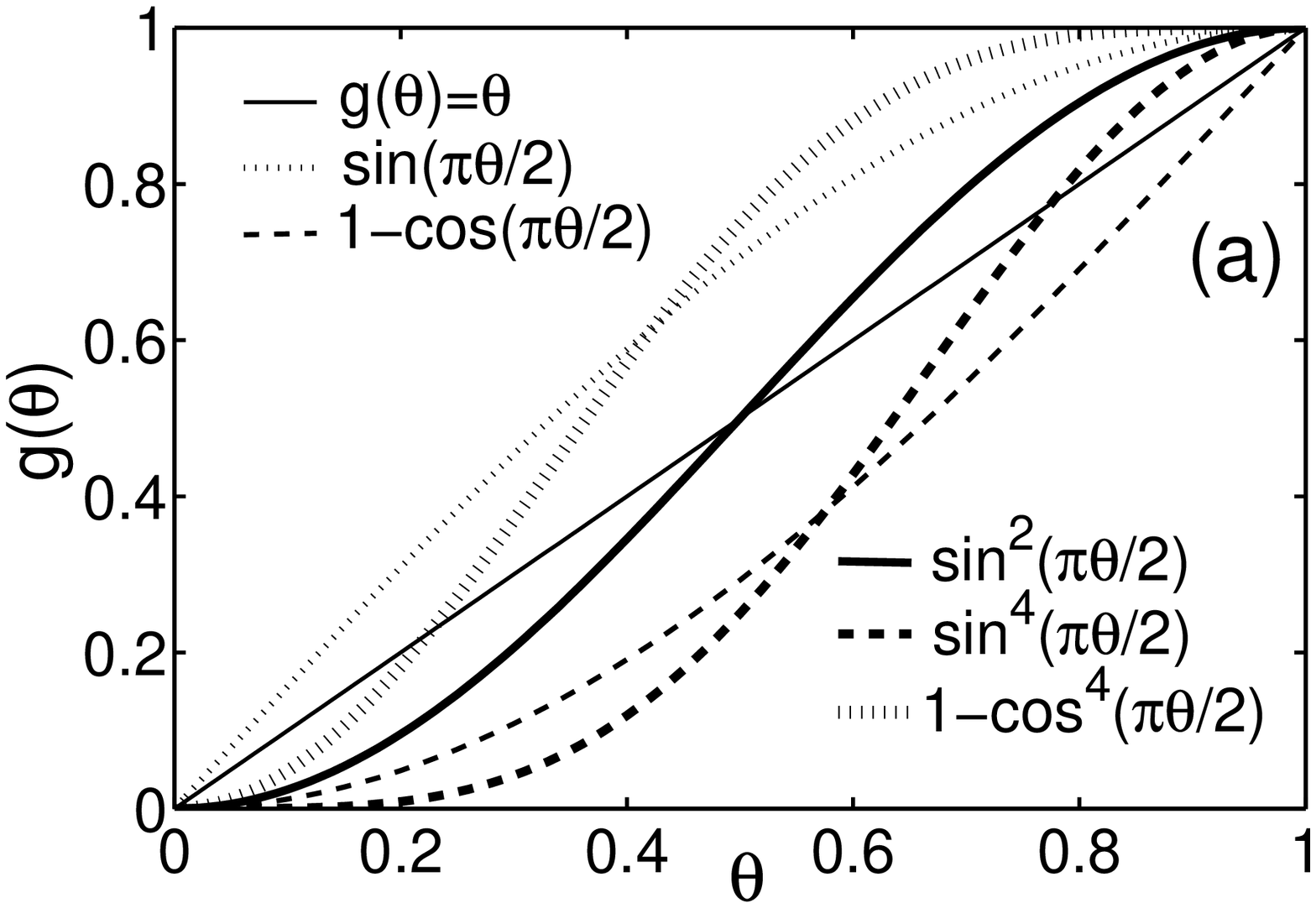}
\includegraphics[width=2.65in]{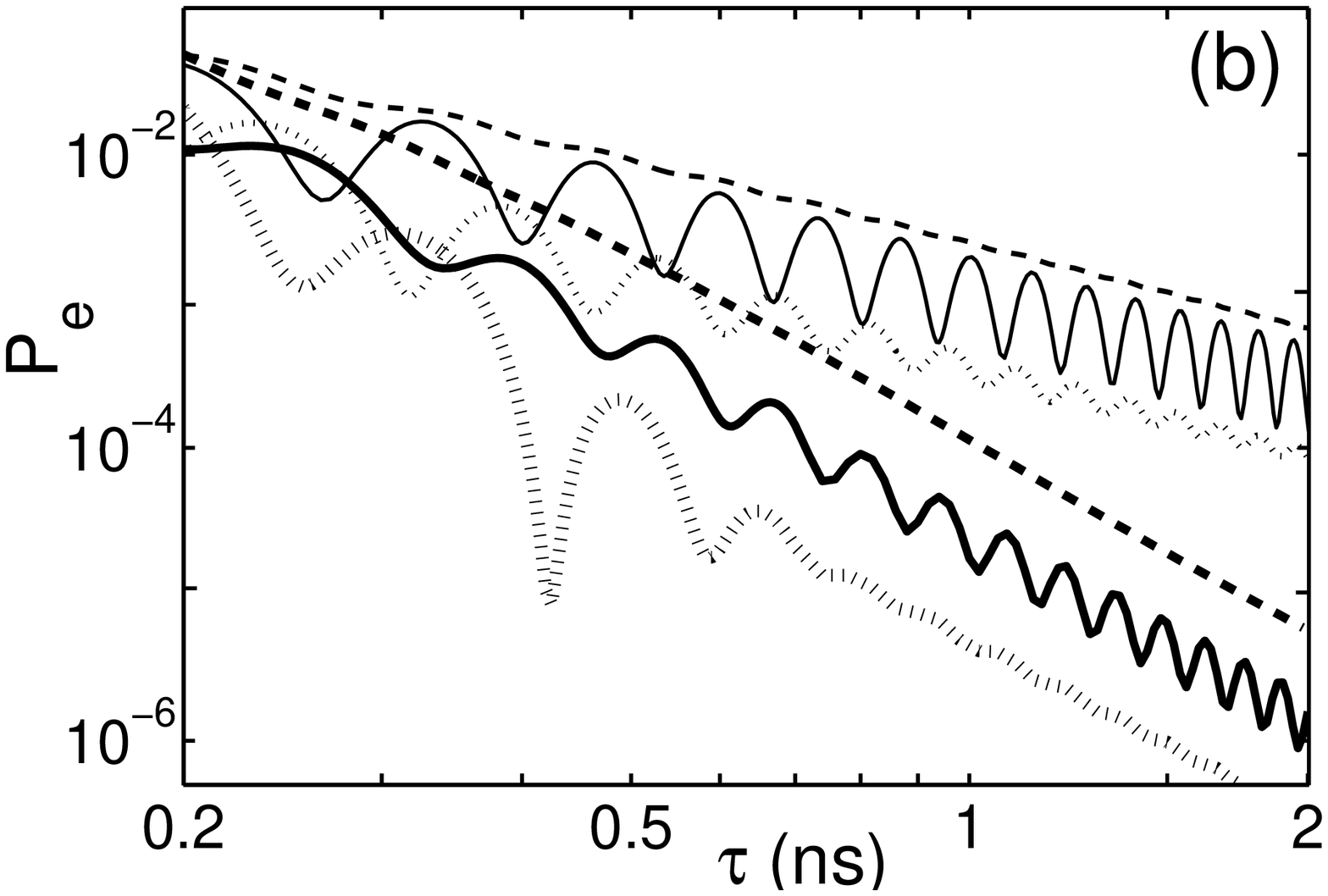}
\caption{(a) Several shapes of the measurement pulse and (b) the
corresponding dependences $P_e(\tau)$.
 The pulse shapes are $g(\theta)=\theta$ (thin solid line),
$\sin(\pi\theta/2)$ (thin dotted line), $1-\cos(\pi\theta/2)$ (thin dashed
line), $\sin^2(\pi\theta/2)$ (thick solid line), $\sin^4(\pi\theta/2)$
(thick dashed line), and $1-\cos^4(\pi\theta/2)$ (thick dotted line).
 }
\label{PE}\end{figure}

The numerically calculated dependence $P_e(\tau)$ for several pulse shapes
[shown in Fig.\ \ref{PE}(a)] is presented in Fig.\ \ref{PE}(b). We consider
the following shapes $g(\theta)$:
 \bea
&&\theta,\ \sin(\pi\theta/2),\ 1-\cos(\pi\theta/2),\ \nonumber\\
&&\sin^2(\pi\theta/2),\ \sin^4(\pi\theta/2),\ 1-\cos^4(\pi\theta/2).
 \ea{4.26}
 The long-$\tau$ behavior of the curves in Fig.\ \ref{PE}(b) can be checked
to satisfy Eq.\ (\ref{4.21}). In particular, at long $\tau$ the error decays
approximately as $\tau^{-2j}$ where $j$ [defined right after Eq.\
(\ref{4.21}) as the lowest order of nonvanishing derivatives at the pulse
endpoints] is equal to 1 or 2 for the shapes shown in the first or second
lines of Eq.\ (\ref{4.26}), respectively (thin and thick lines in Fig.\
\ref{PE}). By designing pulses to be even smoother at the endpoints (having
larger $j$) the decay of $P_e$ with increasing $\tau$ can formally be made
even faster. However, this does not make much practical sense because for
long $\tau$ the error is very small anyway, and because the derivation of
Eq.\ (\ref{4.21}) required an assumption that the pulse shape is described
by an analytical function; this assumption can hardly be applied to an
experimental situation.

From experimental point of view the most interesting range of $P_e$ is
$10^{-4} - 10^{-2}$ (recall that the real measurement error can be up to
$Q_e=\sqrt{P_e}$). We see that in this range $P_e$ still significantly
depends on the shape of the pulse. Therefore, for a fast and reliable
measurement the choice of a proper pulse shape can be really important.
Notice that the oscillating behavior of $P_e(\tau )$ dependence for some
shapes can in principle be used to minimize nonadiabatic measurement error
by choosing specific duration of the measurement pulse.

\section{Discrimination between qubit states by tunneling}

    In this section we neglect the nonadiabatic error discussed above and
consider the measurement error arising during the tunneling stage of the
measurement process. Ideally, the system in state $|1\rangle$ should tunnel
from the left to the right well of the qubit potential (Fig.\ 1), while state
$|0\rangle$ should remain in the left well. However, since the ratio of
tunneling rates $\Gamma_1/\Gamma_0$ for the two states is finite,
discrimination between the states by tunneling is not complete: for a too
short measurement pulse there is a chance that the state $|1\rangle$ remains
in the left well, while for a too long pulse there is a chance that the state
$|0\rangle$ tunnels out.

\begin{figure}
\centering
\includegraphics[width=2.9in]{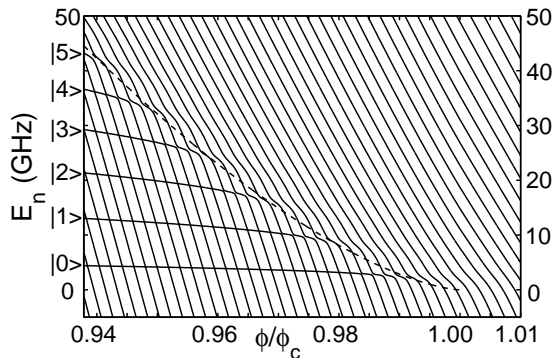}
\caption{Structure of the energy levels $E_n$ (in GHz) as a function of
external flux $\phi$ (in units of critical flux $\phi_c=5.43$). The levels
with $n$ between 168 and 218 (counted from the right-well bottom) are shown.
 The dashed line shows the barrier height $\Delta U_l$.
 The vectors $|k\rangle$ on the left enumerate the states localized in the
 left well.
 }
\label{E-phi}\end{figure}

    To take tunneling into account we consider the exact potential (\ref{2.1})
and still assume qubit parameters of Eq.\ (\ref{2.16}). Figure \ref{E-phi}
shows the dependence of the energy spectrum on the applied flux $\phi$. The
flux is normalized by the critical value $\phi_c$ at which the barrier
between the two wells disappears (for our parameters $\phi_c=5.43$).
 Zero of the energy is chosen at the bottom of the left well for
$\phi<\phi_c$ and at the point of inflection of the potential \cite{pre}
$\delta_c=\pi/2+\arcsin(1/\lambda)$ for $\phi\ge\phi_c$.
 The dashed line shows the barrier height, which satisfies quite
well the cubic-potential approximation \cite{joh05,pre,lik86} $\Delta
U_l\propto(1-\phi/\phi_c)^{3/2}$.
 This line separates states above the barrier which are delocalized
and states under the barrier which are localized either in the left well or
in the right well.
 In Fig.\ \ref{E-phi} almost horizontal branches are the left-well states.
They are denoted on the left side of the frame as $|k\rangle\
(k=0,1,\dots)$, where $k$ enumerates the left-well states.
 The branches with a steep slope below the barrier top correspond to
the right-well states.
 At $\phi>\phi_c$ the potential represents a single well.
 Notice the (anti)crossings between energy levels belonging to
the left and right wells. In the case when the energy relaxation and
dephasing are sufficiently weak so that the corresponding broadening of the
right-well levels is smaller than their energy separation, the level
crossings correspond to enhanced tunneling from the left well.

\begin{figure}
\centering
\includegraphics[width=2.65in]{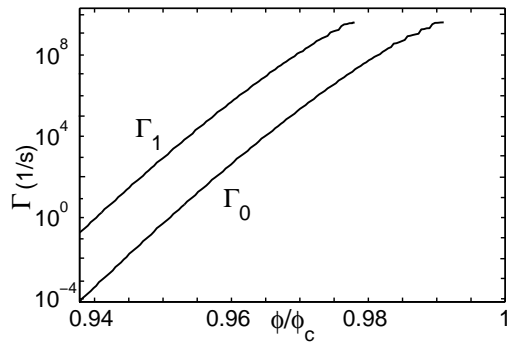}
\caption{The WKB tunneling rates $\Gamma_1$ and $\Gamma_0$ for the states
$|1\rangle$ and $|0\rangle$ as functions of the applied flux $\phi$.
 }
\label{Gamma-fig}\end{figure}

    For simplicity, let us first neglect the level discreteness in the right
well and calculate the tunneling rates $\Gamma_1$ and $\Gamma_0$ for states
$|1\rangle$ and $|0\rangle$ using WKB approximation. Then
  $\Gamma_k = f_a(k) D(k)$,
where WKB factor $D(k)=\exp[-(2/\hbar)\int_{\delta_{t1}}^{\delta_{t2}}
\sqrt{2m(U(\delta)-\tilde{E}_k)}\, d\delta]$ is determined by the integral
between classical turning points $\delta_{t1}$ and $\delta_{t2}$ (we use
notation $\tilde{E}_k$ for the energy of the qubit state $|k\rangle$ to
distinguish it from notation $E_k$ used in Sections II and III). For the
attempt frequency $f_a(k)$ we use approximation \cite{Lik-81}
$f_a(k)=(\omega_l/2\pi)[(2\pi)^{1/2}/k!][(k+1/2)/e]^{k+1/2}$, where
$\omega_l$ is the left-well plasma frequency. (Even though this formula has
been derived \cite{Lik-81} for the case when there are many levels in the
well, and so it is not quite accurate in our case, the inaccuracy is not very
significant for us.) Figure \ref{Gamma-fig} shows the tunneling rates
$\Gamma_1$ and $\Gamma_0$ as functions of $\phi/\phi_c$. The lines end when
the corresponding levels reach the barrier top. Small steps near the ends of
the lines are due to level anticrossings. One can see that the ratio
$\Gamma_1/\Gamma_0$ decreases with increasing flux $\phi$ and is slightly
over $10^2$ in the practically interesting range where $\Gamma_1\agt 10^8$
s$^{-1}$.

\begin{figure}
\centering
\includegraphics[width=2.7in]{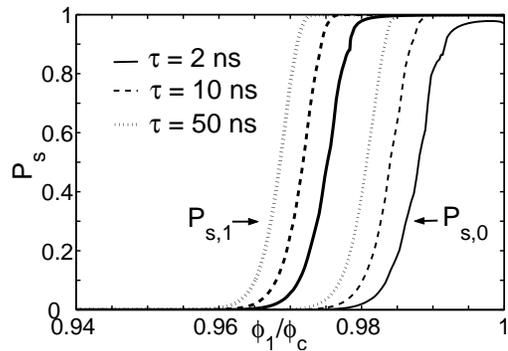}
\caption{The probabilities $P_{s,1}$ (thick lines) and $P_{s,0}$ (thin lines)
of tunneling to the right well during the measurement pulse, starting from
the states $|1\rangle$ or $|0\rangle$, correspondingly, as functions of the
maximum flux $\phi_1$ during the pulse. Calculations are based on the WKB
approximation. Initial flux is $\phi_0/\phi_c=0.94$ ($N=5$). The pulse
duration is $\tau = 2$ ns (solid lines), 10 ns (dashed lines), or 50 ns
(dotted lines).
 }
\label{Scurve-WKB}\end{figure}

Since the applied flux $\phi$ changes during the measurement pulse, the
tunneling should be integrated over the pulse duration. For definiteness let
us consider the parabolic measurement pulse
 \be
\phi(t)=\phi_0+4(\phi_1-\phi_0)(t/\tau)(1-t/\tau)\ \
 (0\le t\le\tau),
 \e{2.2}
so that the flux $\phi(t)$ increases from $\phi_0$ to the maximal value
$\phi_1$ at $t=\tau/2$ and then decreases back to $\phi_0$ at $t=\tau$.
(Notice a different definition of the pulse duration $\tau$ compared to the
definition used in Sec.\ III, in which we considered only half-pulses.)
Figure \ref{Scurve-WKB} shows the probabilities of tunneling during the
measurement pulse (switching to the right well) $P_{s,1}$ and $P_{s,0}$ for
levels $|1\rangle$ and $|0\rangle$,
  \be
  P_{s,k} = 1- \exp \left[ -\int_0^\tau \Gamma_k(t)\, dt \right] ,
  \e{P_si}
as functions of the maximum flux $\phi_1$ (normalized by critical flux
$\phi_c=5.43$); initial flux $\phi_0=5.09$ ($\phi_0/\phi_c=0.94$) corresponds
to $N=5$; the pulse durations are $\tau=2$ ns, 10 ns, and 50 ns.  Notice that
$\Gamma_k$ is not well-defined when the state goes over the barrier, while
we need some value of $\Gamma_k$ for integration in Eq.\ (\ref{P_si}); in
this case we still use the definition $\Gamma_k = f_a(k) D(k)$ but assume
$D(k)=1$.
   The curves in Fig.\ \ref{Scurve-WKB} remain at practically zero level
for small pulse amplitudes and saturate at 100\% level for large enough
pulse amplitude (the ``S-curve'' shape). (The switching probability $P_{s,0}$
for $\tau=2$ ns does not fully approach 100\% because the pulse duration is
too short.) The most important observation is that the flux shift between
the curves for $P_{s,1}$ and $P_{s,0}$ (for the same $\tau$) is sufficiently
large to reliably distinguish states $|1\rangle$ and $|0\rangle$. However,
the measurement error, which can be defined as $(P_{s,0}+1-P_{s,1})/2$, is
finite for any $\phi_1$, and in the optimal point is on the order of
$10^{-2}$.  (Actually, our method is not quite accurate for the top part of
the S-curves because of significant level anticrossings, which leads to
visible kinks on the curves in Fig.\ \ref{Scurve-WKB} and inaccuracy of the
measurement error calculation.) One can see that the minimal measurement
error improves (decreases) with increase of the pulse duration $\tau$; even
though the separation of the two S-curves decreases with increase of $\tau$,
they become sharper, leading to better discrimination between states
$|1\rangle$ and $|0\rangle$.

    The WKB approximation does not take into account the level discreteness
in the right well. To take it into account, we have performed the full
quantum-mechanical simulation of the evolution (\ref{4.2})--(\ref{4.15})
during the measurement pulse (\ref{2.2}). Since the level broadening is
determined by energy relaxation (and dephasing), which is very difficult to
simulate exactly, we have used a simpler model. We assume that the levels in
the right well and above the barrier have a finite lifetime (complex energy),
while levels in the left well do not decay. The decay rate for the $n$th
level counting from the bottom of the right well is chosen as
$\gamma=n/T_1$, where $T_1$ is the energy relaxation time; the corresponding
width of the broadened level is $\hbar\gamma$. Since there are many levels
in the right well and we are interested only in levels close to the barrier
top, the level index $n$ can be replaced with the number of levels $N_r$ in
the right well; for the parameters (\ref{2.16}) which we use, $N_r=174$ at
$N=5$. [Actually, there are some additional details in the algorithm. For
example, to prevent unphysical decay of the left-well state during the level
crossing, we smoothly (parabolically) suppress $\gamma$ when the energy
distance to the nearest level becomes less than $1/5$ of the normal level
separation in the right well. The numerical criterion for a left-well level
is that its energy is in between the left well bottom and the barrier top,
and also the maximum of the wavefunction is to the left of the barrier top.]

\begin{figure}
\centering
\includegraphics[width=2.9in]{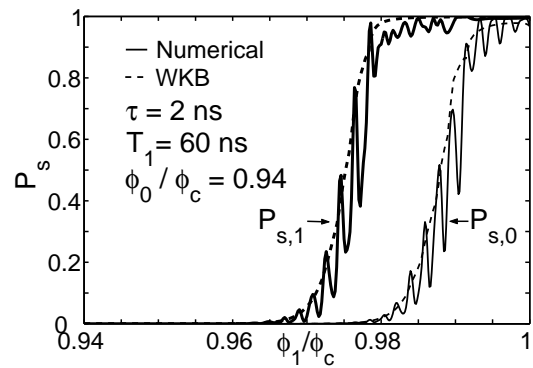}
\caption{The switching probabilities $P_{s,0}$ (thin lines) and $P_{s,1}$
(thick lines) as functions of the maximum flux $\phi_1$, calculated
numerically (solid lines) and using WKB approximation (dashed lines). The
measurement pulse is of parabolic shape with duration $\tau=2$ ns, starting
and ending at flux $\phi_0/\phi_c=0.94$ ($N=5$). Oscillations of the solid
lines reflect level discreteness in the right well. (At $N=5$ there are
$N_r=174$ levels in the right well.)
 }
\label{Scurve-discr}\end{figure}

 Solid lines in Fig.\ \ref{Scurve-discr} show  the numerically calculated
probability $P_{s,k}$ of switching to the right well during the measurement
pulse (starting with $N=5$, $\phi/\phi_c=0.94$) in the case when initially
either the left-well ground state ($k=0$) or the first excited state ($k=1$)
is occupied. Assumed pulse duration is $\tau=2$ ns, while $T_1=60$ ns.
Switching probability is defined as $P_s=1-P_l$, where $P_l$ is the total
population of the left-well levels after the pulse. One can see that the
numerical dependence $P_{s,k}(\phi )$ is generally similar to the
corresponding WKB result (dashed lines); however, the numerical curves show
significant oscillations due to level discreteness in the right well. Notice
that the amplitude of oscillations in this case is mainly determined not by
the width of the levels (which is still relatively small in this example),
but by continuous sweeping through the comb of levels during the pulse. (For
our parameters the level broadening becomes larger than the level spacing at
$T_1\alt 5$ ns.) Similar oscillations in the dependence $P_{s,k}(\phi_1)$
have been observed experimentally, \cite{Katz-tbp} though for a qubit with a
smaller number of levels in the right well ($N_r\sim 30$, in contrast to
$N_r \sim 170$ in the example shown in Fig.\ \ref{Scurve-discr}), that leads
to a smaller level broadening $\hbar\gamma \simeq \hbar N_r/T_1$ and
therefore more pronounced effects of the right-well level discreteness.

    Figures \ref{Gamma-fig}--\ref{Scurve-discr} show that the quality of
tunneling discrimination between states $|1\rangle$ and $|0\rangle$ depends
significantly on the choice of the measurement pulse amplitude (maximum flux
$\phi_1$), and at the optimal point the measurement error
$(P_{s,0}+1-P_{s,1})/2$ is on the order of $10^{-2}$. This error decreases
for longer measurement pulses and can be further reduced by changing the
qubit parameters.

\section{Left well repopulation}
\label{IV}

A steady progress in fabrication of phase qubits has resulted in a
significant decrease of the dissipation rate.\cite{mar05} This trend will
obviously continue because of the requirement of very low decoherence in
quantum computation. Therefore it is of interest to study the measurement
process for a phase qubit with a very long energy relaxation time $T_1$. Let
us we consider the limit of negligible dissipation, $T_1= \infty$. As we
will see below, in this case the results are qualitatively different from the
results discussed in the previous section. The most important effect is that
the switching probability $P_{s,k}$ does not approach 100\% even for
measurement pulses with maximum flux $\phi_1$ well above the critical value
$\phi_c$, in contrast to what is shown in Figs.\ \ref{Scurve-WKB} and
\ref{Scurve-discr}.

\begin{figure*}[htb]
\includegraphics[width=12.5cm]{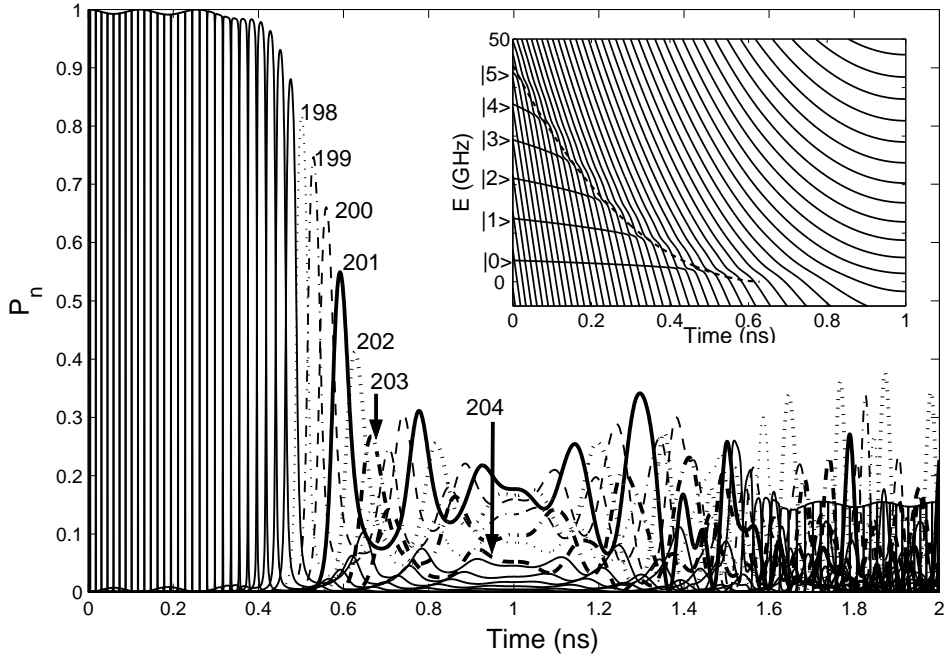}
\caption{The level populations $P_n$ ($169\le n\le204$) as functions of time
$t$ for the measurement pulse with duration $\tau=2$ ns, starting and ending
at $\phi_0/\phi_c=0.94$ ($N=5$) with maximum flux in the middle
$\phi_1/\phi_c=1.01$.
 The curves for $169\le n\le197$ are thin solid lines, the
curves for $198\le n\le204$ are denoted by the value of $n$.
 Inset: Energies $E_n\ (168\le n\le218)$ versus time $t$ for the rising
part of the pulse (similar to Fig.\ \protect\ref{E-phi}).
 }
 \label{f7}\end{figure*}

To demonstrate this effect, let us assume that initially the system is in
the left-well ground state $|0\rangle$ and simulate the full quantum
evolution (\ref{4.2})--(\ref{4.15}) during the measurement pulse
(\ref{2.2}). Figure \ref{f7} (the main frame) shows the populations $P_n(t)$
of levels $n$ (levels are counted from the right-well ground state) with
$169\le n\le204$ for the measurement pulse with initial flux
$\phi_0=0.94\phi_c$ (corresponding to $N=5$), maximum flux
$\phi_1=1.01\phi_c$, and total duration $\tau=2$ ns. The inset shows the
structure of energy levels $E_n$ (for $168\le n\le 218$) as a function of
time during the first (increasing) half of the pulse (the only difference
between Fig.\ \ref{E-phi} and the inset of Fig.\ \ref{f7} is the converted
horizontal axis). The state $|0\rangle$ initially corresponds to $n=169$.

 The population of the qubit state $|0\rangle$  can be seen as the upper
envelope dependence of $P_n(t)$ for $t<0.63$ ns; similarly, lower envelope
curve at $t<0.4$ ns corresponds to the state $|1\rangle$. At $t< 0.4$ ns the
state $|0\rangle$  is sufficiently deep in the left well, so that the level
splittings in the avoided crossings are too small to cause transitions from
this level to the right well (notice though the change of the level number
corresponding to $|0\rangle$ after each level crossing).
 As a result, at $t<0.4$ ns the populations of levels
$|0\rangle$ and $|1\rangle$ vary only due to the nonadiabatic effects (see
Sec.\ \ref{III}).

  At $0.4<t<0.63$ ns there is an appreciable probability
of a transition to the right well at each crossing of the qubit level
$|0\rangle$ with a right-well level.
 In this time interval one can still interpret the envelope of the
populations of the crossing states as the population of state $|0\rangle$,
despite the level $|0\rangle$ becomes above the barrier at $t>0.5$ ns (see
the inset).
 We have checked  numerically  that the transition probabilities at
the level crossings agree with the Landau-Zener formula \cite{lan77}.
 At $t<0.5$ ns, when the state $|0\rangle$ is still below the barrier,
the transitions to the right well are due to tunneling.
 At $t>0.5$ ns the level splittings increase and become comparable
to the separations between adjacent right-well levels; this means that
generally more than two levels are coupled simultaneously.
 At $t > 0.63$ ns the left well disappears (see the inset), and the
initial population of level $|0\rangle$ becomes distributed between
several states in the resulting single well.

\begin{figure}
\centering
\includegraphics[width=2.8in]{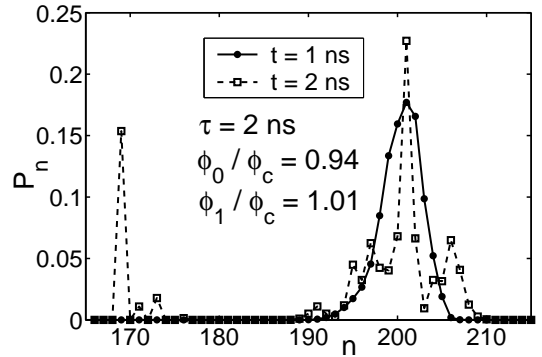}
\caption{The level populations $P_n$ at $t=1$ ns and $t=2$ ns for the
process shown in Fig.\ \protect\ref{f7} ($\phi_1/\phi_c=1.01$, $\tau=2$ ns).
 The data are shown by dots and squares, connected by lines as guides for the eye.
 }
\label{f9}\end{figure}

During the time interval when the potential (\ref{2.1}) has a single-well
shape (0.63 ns $< t<$ 1.37 ns), one can see a rather significant and
complicated redistribution of the level population.
 Notice that the most populated levels in the single well are those
with $198\le n\le 204$, i.e., with the quantum number significantly higher
than the initial $n=169$. The solid line in
 Fig.\ \ref{f9} shows the population distribution at $t=1$ ns (middle of
the pulse); one can see that the population
  is noticeable only for $n\ge192$, which are the states delocalized
(above-the-barrier) before and after the pulse.

 After $t=1$ ns the pulse starts to decrease, and the
potential (\ref{2.1}) starts to return to its initial shape. The levels
localized in the left well start to appear again and cross with the
populated right-well levels which are moving up.
 As a result, the qubit state $|0\rangle$ partially recovers  its
population (in Fig.\ \ref{f7} it is seen as an approximately horizontal line
after 1.6 ns), and also a number of higher levels in the left well become
populated.
 As the dashed line in Fig.\ \ref{f9} shows, after the end of the pulse an
appreciable population is acquired by states $|0\rangle$, $|1\rangle$, and
$|2\rangle$ of the left well and also by a number of delocalized states with
$n\ge190$.

 In short, during the first half of the pulse, right-well and
delocalized levels go down in energy and cross with left-well levels, thus
gaining population, while during the second half of the pulse, right-well
levels move up and cross again with left-well levels, resulting in a partial
repopulation of the initial state and an excitation of higher levels.
Naively, one could expect that after such strong measurement pulse the
population should be transferred from the left well to the right well.
However, we see that without dissipation the population actually goes mainly
to highly excited delocalized states. Assuming a very small but finite
dissipation after the pulse, we conclude that a populated delocalized state
will eventually end up either in the left or in the right well. Therefore,
there are two contributions to the left well repopulation: direct
repopulation (due to level crossings) and relaxation from delocalized states.

After the pulse the probability for the system to be to the right of the
barrier top is
 \be
P_r(t)=\int_{\delta_m}^\infty|\Psi(\delta,t)|^2d\delta
 \e{4.33}
(where $\delta_m$ is the barrier-top position), and generally depends on time
due to interference of occupied delocalized levels.
 We define the probability of escape (switching) to the right well
as the time average of $P_r(t)$.
 Thus, the switching probability is given by
 \be
P_s=\sum_{n}P_{n}\int_{\delta_m}^\infty|\psi_{n}(\delta)|^2d\delta,
 \e{4.32}
where $P_n$ is the population of level $n$ after the pulse, and the integral
is the probability for the system in state $n$ to be in the right well.
 This integral equals zero (one) for states localized in the left
(right) well and is between 0 and 1 for delocalized states.

\begin{figure}
\centering
\includegraphics[width=2.8in]{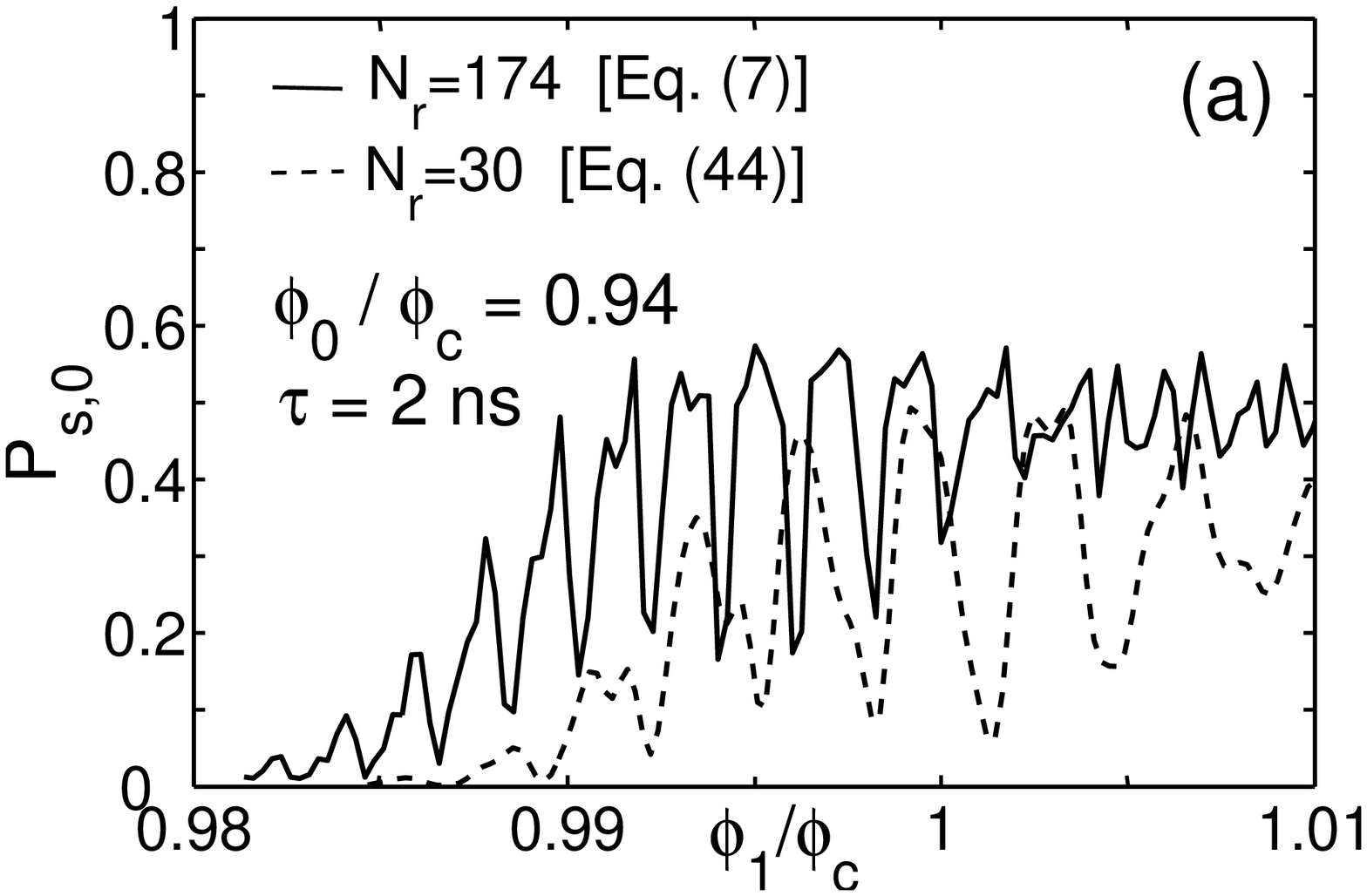}
\includegraphics[width=2.8in]{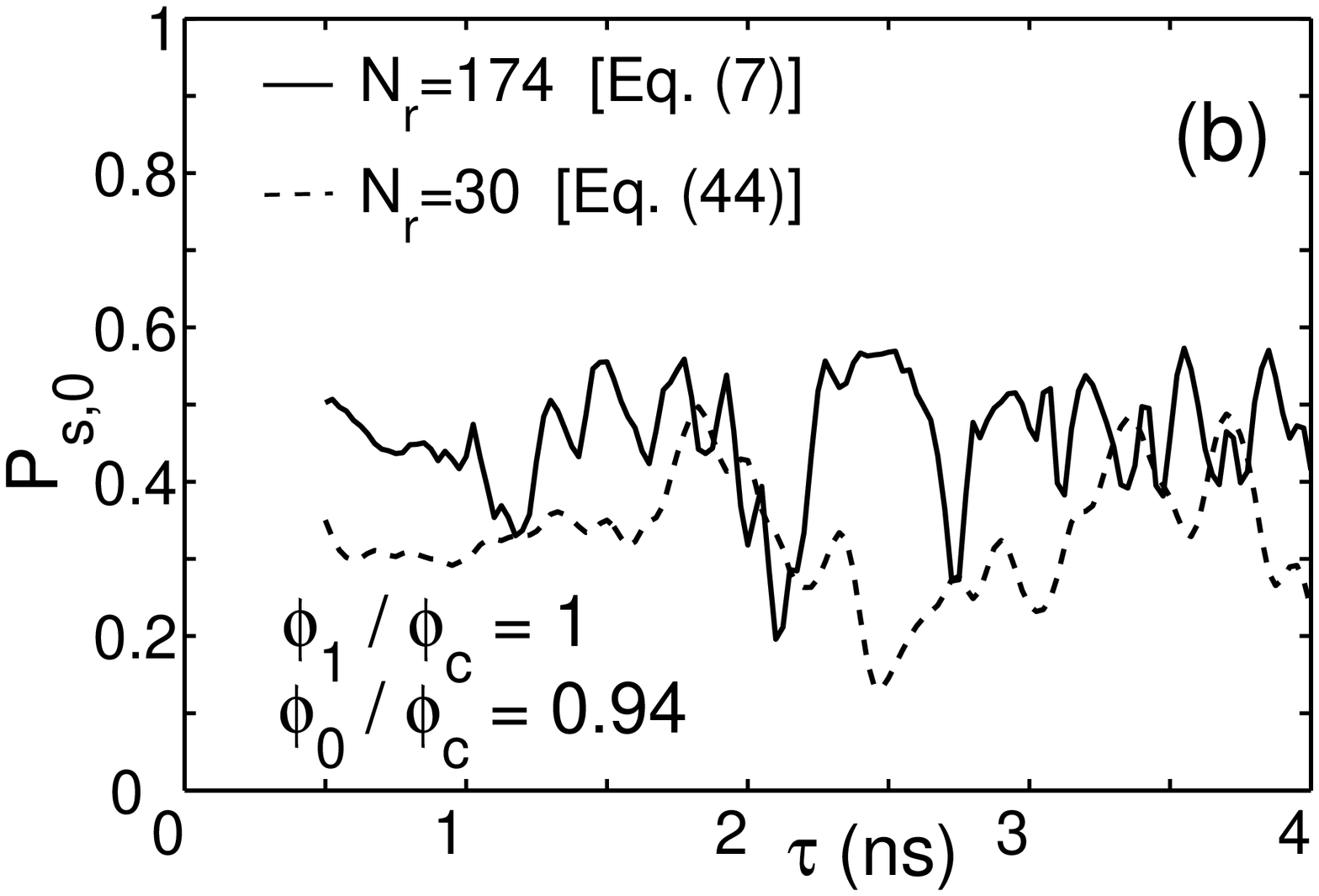}
\caption{The switching probability $P_{s,0}$ for the initial state
$|0\rangle$ [see Eq.\ (\protect\ref{4.32})] (a) versus the maximum flux
$\phi_1$ (normalized by the critical flux $\phi_c$) for $\tau=2$ ns and
 (b) versus the pulse duration $\tau$ for $\phi_1=\phi_c$. Solid lines are
for our usual set (\protect\ref{2.16}) of qubit parameters; dashed lines are
for the qubit with parameters (\protect\ref{4.40}), having much smaller
number $N_r$ of discrete levels in the right well.
 }
\label{f8}\end{figure}

The solid line in Fig.\ \ref{f8}(a) shows the dependence of the switching
probability $P_{s,0}$ (starting from state $|0\rangle$) on the maximum flux
$\phi_1$ for $\tau=2$ ns and $\phi_0/\phi_c=0.94$. The switching probability
practically vanishes at $\phi_1/\phi_c<0.98$ when the initial state
$|0\rangle$ remains sufficiently deep inside the left well and cannot tunnel
(similar to the results shown in Figs.\ \ref{Scurve-WKB} and
\ref{Scurve-discr}). However, in contrast to Figs.\ \ref{Scurve-WKB} and
\ref{Scurve-discr}, an increase of the pulse amplitude leads to significant
and irregular oscillations of $P_{s,1}(\phi_1)$. Most importantly, the
switching probability remains to be significantly less than 100\% even for
$\phi_1$ exceeding $\phi_c$.

    The solid line in Fig.\ \ref{f8}(b) shows the switching probability $P_{s,0}$
as a function of the pulse duration $\tau$ for $\phi_1=\phi_c$. The
dependence is even more irregular than in Fig.\ \ref{f8}(a). The irregular
oscillations seen in Figs.\ \ref{f8}(a) and \ref{f8}(b) can be related to
the complicated time dependence of the level populations seen in Fig.\
\ref{f7}.

\begin{figure}
\centering
\includegraphics[width=2.9in]{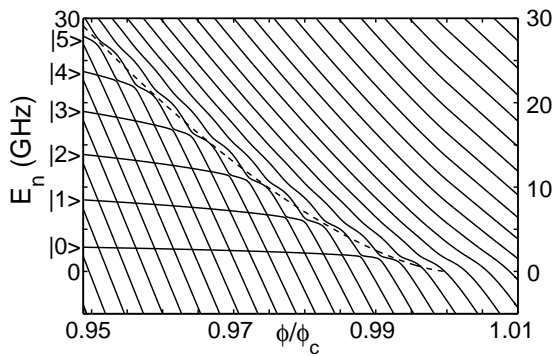}
\caption{Energies $E_n$ (in GHz) as a function of flux $\phi$ (in units of
$\phi_c=3.55$) for the qubit with parameters (\protect\ref{4.40})
corresponding to $N_r=30$.
 The levels with $24\le n\le57$ are shown.
 The dashed line corresponds to the barrier height $\Delta U_l$.
 }
\label{SP}\end{figure}

    Besides calculations for the qubit parameters shown in Eq.\ (\ref{2.16}),
we have also considered a different set of qubit parameters:
   \be
C=790\ \mbox{fF},\ L=0.720\ \mbox{nH},\ I_0=0.764\ \mu\mbox{A}.
 \e{4.40}
 The main difference between the two sets is a significantly different
number of levels in the right well. For the initial condition $N=5$, the
number of levels $N_r$ in the right well is $N_r=174$ for our main set of
parameters (\ref{2.16}) and $N_r=30$ for the parameters (\ref{4.40}). The
eigenenergies of the Hamiltonian as functions of flux $\phi$ for the
parameter set (\ref{4.40}) are shown in Fig.\ \ref{SP} (flux is normalized
by the critical flux $\phi_c=3.55$).

    In spite of a significant difference of the energy level structure, the
behavior of the switching probability $P_{s,0}$ shown in Figs.\ \ref{f8}(a)
and \ref{f8}(b) by dashed lines for the parameter set (\ref{4.40}) is
qualitatively similar to the previously discussed behavior for the set
(\ref{2.16}) (solid lines). Notice, however, that in a real experiment the
effective time $\gamma^{-1}\simeq T_1/N_r$ spent on a right-well level before
relaxation is significantly larger for the set (\ref{4.40}) and therefore
the discussed here effects due to insufficiently fast relaxation will be
more pronounced.

    It is important to notice that while the large irregular oscillations
of switching probability $P_s$ in Fig.\ \ref{f8} is a result of quantum
interference and therefore will be significantly suppressed by a weak
relaxation, the fact that $P_s$ does not approach 100\% even for $\phi_1
>\phi_c$ [as for curves in Fig.\ \ref{f8}(a)] is a much more robust,
essentially classical effect. Our results show that a very significant
repopulation of the left well may occur if the dissipation is insufficiently
fast. In other words, sufficiently strong energy dissipation is a necessary
requirement for a good measurement fidelity. Conversely, the repopulation
effect imposes a lower limit on the pulse duration for a given dissipation
rate.

Let us estimate the condition necessary to eliminate the left-well
repopulation. For simplicity, we assume the pulse to be almost rectangular,
with the rise and fall times being significantly shorter than the time
$\tau_m$, during which the pulse is at its maximum. Let us denote by $n_i$
the level number (counting from the right-well ground state), corresponding
to the right-well level close to the top of the barrier for maximum flux
bias. Assume that this level is populated immediately after the pulse rise.
Then during $\tau_m$ the energy relaxation will lead to the decrease of
$n$-number corresponding to the mostly populated level. Neglecting the
``energy diffusion'' of the populated levels, we estimate the $n$-number at
time $\tau_m$ as $n_i'\simeq  n_i-\gamma \tau_m$, where $\gamma$ is the
transition rate between adjacent levels. We neglect the weak dependence of
$\gamma$ on the level number. Let $n_f$ be the first level above the barrier
top after the pulse is over. If we neglect a change of $n_i'$ during the
falling part of the pulse, the condition that there is no return to the left
well is $n_i'<n_f$, which means $\tau_m>(n_i-n_f)/\gamma$. Using an estimate
$\gamma\simeq n_i/T_1$, we obtain
 \be
 \tau_m>\frac{n_i-n_f}{n_i}\, T_1.
 \e{1}
This condition is actually a lower bound; the required pulse duration
$\tau_m$ can be even longer because of not immediate population of the level
$n_i$, widening of the range of populated levels in the course of relaxation
(``energy diffusion''), and possibility of repopulation due to tunneling.
 For the case shown in Figs.\ \ref{f7} and \ref{f9} we find $n_i=205$,
 $n_f=181$, yielding $\tau_m>0.13\,T_1$.

    The necessity of sufficiently long measurement pulses thus seems to be
an important requirement for the design of quantum gates based on phase
qubits. It also makes harder the solution of the cross-talk
problem\cite{mcd05,pre} for simultaneous measurement of several qubits,
which argues in favor of designing adjustable coupling between qubits.

\section{Conclusion}
\label{V}

In this paper we have studied the behavior of a flux-biased phase qubit in
the process of its measurement (using analytical approaches and numerical
solution of the Schr\"{o}dinger equation) and analyzed several mechanisms
leading to measurement errors.

First, we have studied nonadiabatic errors (Sec.\ III), which occur during
the rising part of the measurement pulse (before the tunneling stage) due to
finite duration of the pulse, leading to the transitions between qubit
states $|1\rangle$ and $|0\rangle$. We have developed simplified analytical
approaches with several levels of accuracy [see Eqs.\ (\ref{Exact}),
(\ref{Simple}), and (\ref{simple-simple})] and compared them with the
numerical results (Sec.\ \ref{IIIB}). The nonadiabatic error generally
decreases with increase of the pulse duration $\tau$; however, this
dependence may exhibit significant oscillations. The numerical value of the
error depends significantly on the pulse shape (Figs.\ \ref{comp-simple} and
\ref{PE}), thus showing importance of a proper pulse shape design. For pulse
duration over few nanoseconds the nonadiabatic error is typically
sufficiently small to be considered negligible.

   Another type of measurement error is due to incomplete discrimination
between states $|1\rangle$ and $|0\rangle$ during the tunneling stage of the
measurement process (Sec.\ IV). For typical qubit parameters\cite{mcd05} the
ratio of WKB tunneling rates $\Gamma_1/\Gamma_0$ for states $|1\rangle$ and
$|0\rangle$ is few times $10^2$ and decreases with the applied flux bias
(i.e.\ when faster measurement is required). Therefore, the minimized (over
the maximum applied flux during measurement) measurement error is smaller
for longer pulses (Fig.\ \ref{Scurve-WKB}). The level discreteness in the
right well of the qubit potential [Fig.\ \ref{Schematic}(b)] leads to
oscillations in the dependence of the switching probability on the
measurement pulse amplitude (Fig.\ \ref{Scurve-discr}). A typical value of
the minimized error due to incomplete discrimination between the states by
tunneling is on the order of $10^{-2}$ for typical qubit parameters of
present-day experiments.\cite{mcd05} Change of qubit design may lead to
further reduction of this type of measurement error.

    In Sec.\ V we have analyzed the quantum evolution of the qubit during its
measurement in the case of complete absence of energy relaxation, and found
that the energy relaxation plays a very important role in the process of
qubit measurement. In the case of insufficiently fast dissipation
(characterized by energy relaxation time $T_1$) the measurement error can be
caused by the repopulation of the left well of the qubit potential [Fig.\
\ref{Schematic}(b)] after the tunneling stage of the measurement process. A
simple estimate (\ref{1}) for the repopulation effect shows that reliable
measurement requires relatively long measurement pulses; in the analyzed
example the pulse should be longer than $\sim 0.1\, T_1$. This result may be
quite important for the design of the quantum gates based on phase qubits.
Some alleviation of the problem may be achieved by using a different shape
of the measurement pulse, so that the barrier height of the qubit potential
after the pulse is significantly lower than before the measurement.

  The work was supported by NSA and DTO under ARO grant W911NF-04-1-0204.


\begin{thebibliography}{99}
\bibitem{N-difference} We use the same value $N=1.355$ as
in our previous paper \protect\cite{pre} for the two-qubit setup. Since in
the present paper the effective junction capacitance is a little smaller
(because we do not have coupling capacitance $C_x=6$ fF), the state
$|1\rangle$ for $N=1.355$ is actually slightly above the barrier for the
potential (\protect\ref{2.1}). However, for the modified potential
(\protect\ref{4.8}) used in Section III the state $|1\rangle$ is still below
the barrier for $N=1.355$.
\end{thebibliography}
\end{document}